%% file: main.tex
\documentclass[twoside,11pt]{article}

\usepackage{lastpage}
\usepackage[accepted,specialissue]{melba}

\input{Macros}

\usepackage{amsmath,amsfonts}
\usepackage{multirow, makecell}
\usepackage{booktabs}
\usepackage{siunitx}
\usepackage{verbatim}
\usepackage{algorithm}
\usepackage{algpseudocode}
\usepackage{xcolor}
\usepackage{bbding}
\usepackage{utfsym}
\usepackage{txfonts}
\usepackage{fontawesome}

\usepackage{color}
\usepackage{colortbl}
\usepackage{multirow}
\usepackage{tablefootnote}
\usepackage{tikz}
\usetikzlibrary{arrows.meta}
\usepackage[T1]{fontenc}

\newcommand{\blue}[1]{#1}

\melbaid{2024:001}  
\melbaauthors{Singh, Denker, Barbano, Kereta, Jin, Thielemans, Maass, and Arridge}  
\volume{2}
\firstpageno{547} 
\melbayear{2024}  
\datesubmitted{08/2023}  
\datepublished{01/2024}  

\melbaspecialissue{Special Issue for Generative Models}
\melbaspecialissueeditors{Mert Sabuncu, Sotirios A. Tsaftaris}
\doi{https://doi.org/10.59275/j.melba.2024-5d51}

\ShortHeadings{Score-Based Generative Models for PET Image Reconstruction}{Singh, Denker, Barbano, Kereta, Jin, Thielemans, Maass and Arridge}

\title{Score-Based Generative Models for PET Image Reconstruction}

\author{\firstname Imraj R D \surname Singh\thanks{Equal contribution.}~\orcid{0000-0003-2186-0977} \email imraj.singh.20@ucl.ac.uk \\  
	\addr Department of Computer Science,  University College London, 66-72 Gower St, WC1E 6EA, London, United Kingdom.
	\AND
	\firstname Alexander \surname Denker\footnotemark[1]~\orcid{0000-0002-7265-261X} \email adenker@uni-bremen.de \\
	\addr Center for Industrial Mathematics, University of Bremen, Bibliothekstr. 5, 28359 Bremen, Germany. 
    \AND 
    \firstname Riccardo \surname Barbano\footnotemark[1]~\orcid{0000-0003-1863-2092} \email riccardo.barbano.19@ucl.ac.uk\\
    \addr Department of Computer Science,  University College London, 66-72 Gower St, WC1E 6EA, London, United Kingdom.
    \AND
    \firstname \v{Z}eljko \surname Kereta~\orcid{0000-0003-2805-0037}  \email z.kereta@ucl.ac.uk \\ 
    \addr Department of Computer Science,  University College London, 66-72 Gower St, WC1E 6EA, London, United Kingdom.
    \AND 
    \firstname Bangti \surname Jin~\orcid{0000-0002-3775-9155} \email b.jin@cuhk.edu.hk\\ 
    \addr Department of Mathematics, The Chinese University of Hong Kong, Shatin, N.T., Hong Kong.
    \AND
    \firstname Kris \surname Thielemans~\orcid{0000-0002-5514-199X} \email k.thielemans@ucl.ac.uk\\ 
    \addr Institute of Nuclear Medicine, University College London, London, United Kingdom.
    \AND 
    \firstname Peter \surname Maass~\orcid{0000-0003-1448-8345} \email pmaass@uni-bremen.de \\ 
    \addr Center for Industrial Mathematics, University of Bremen, Bibliothekstr. 5, 28359 Bremen, Germany.
    \AND
    \firstname Simon \surname Arridge~\orcid{0000-0003-1292-0210} \email s.arridge@ucl.ac.uk\\
    \addr Department of Computer Science,  University College London, 66-72 Gower St, WC1E 6EA, London, United Kingdom.
}

\begin{document}

\maketitle

\begin{abstract}
Score-based generative models have demonstrated highly promising results for medical image reconstruction tasks in magnetic resonance imaging or computed tomography. However, their application to Positron Emission Tomography (PET) is still largely unexplored. PET image reconstruction involves a variety of challenges, including Poisson noise with high variance and a wide dynamic range. To address these challenges, we propose several PET-specific adaptations of score-based generative models. The proposed framework is developed for both 2D and 3D PET. In addition, we provide an extension to guided reconstruction using magnetic resonance images. We validate the approach through extensive 2D and 3D \textit{in-silico} experiments with a model trained on patient-realistic data without lesions, and evaluate on data without lesions as well as out-of-distribution data with lesions. This demonstrates the proposed method's robustness and significant potential for improved PET reconstruction.
\end{abstract}

\section{Introduction}
Positron Emission Tomography (PET) is a functional medical imaging technique for quantifying and visualising the distribution of a radio-tracer within the body, and is vital in clinical practice for accurate diagnosis, treatment planning, and monitoring of diseases. 
In a PET scan, radio-tracers are injected to probe a specific biological pathway of interest. 
Through the decay of the radio-tracer a positron is released, which upon annihilating with an electron produces a pair of coincident photons that travel in approximately anti-parallel directions.
These emitted photons are detected and are then used to reconstruct the underlying radio-tracer distribution. 
The relationship between the measured emissions and the radio-tracer can be approximated with the Poisson noise model as
\begin{equation}
    \label{eq:pet_inverse_problem}
        \By \sim \cP(\Bbary), \quad \Bbary = \BA \Bx + \Bbarb,
\end{equation}
where $\Bbary\in\mathbb{R}^m$ is the expected value of the measurements ($m$ is the number of detector bins) and~\mbox{$\Bx\in\mathbb{R}^n$} is the discrete (voxel) basis representation of the tracer distribution ($n$ is the number of voxels). 
The system matrix $\BA\in\mathbb{R}^{m\times n}$ includes approximate line integrals between detectors as well as physical phenomena such as photon attenuation, positron range, and detector sensitivity. It should be noted that 3D measurements detect pairs of photons between detector rings, i.e. they are not a stack of 2D measurements. The expected background $\Bbarb\in\mathbb{R}^m$ are estimates of scatter and randoms events \cite{QiLeahy:2006}. 
The unique challenges that distinguish PET from other imaging modalities, e.g. Magnetic Resonance Imaging (MRI) and Computed Tomography (CT), include Poisson noise with low mean number of counts, and widely varying dynamic range of images due to functional differences between patients. 

Most inverse problems in imaging are ill-posed, in the sense that the solution may not exist, not be unique, or not depend continuously on the measurement noise \citep{EnglHanke:1996,ItoJin:2015}.
To stabilise the reconstruction process, prior knowledge is often leveraged through a penalising functional that promotes solutions from a desirable image subset.
The priors are typically hand-crafted to promote desired features in the reconstructed image, such as sparsity of edges \citep{Rudin1992} or smoothness. 
Furthermore, 
if an additional image is available, e.g. with high resolution structural information, a suitable prior can promote common features between the two images, commonly referred to as guided reconstruction \citep{Ehrhardt2021}. 
In recent years, deep learning approaches have shown state-of-the-art performance in PET image reconstruction, see surveys \citep{Reader2020DLPET,Pain2022DLPET}. 
Existing approaches include post-processing \citep{Kaplan2018}, to synthesise high-dose images from low-dose ones (which is akin to denoising), and deep unrolled optimisation \citep{Abolfazl2021,guazzo2021learned}. 
However, these supervised approaches require large volumes of paired data that is often hard to acquire, and their performance may degrade greatly in the presence of distributional shift  \citep{antun2020instabilities,darestani2022test}.

In contrast, generative models 
require a dataset only of images of the target domain.
These can, for example, be high-quality reconstructions acquired from prior scans. The aim of generative modelling is to approximate the image manifold of a given dataset \citep{bengio2013representation}. 
There are a variety of methods for this task, e.g. generative adversarial networks \citep{goodfellow2014generative}, variational autoencoders \citep{kingma2013auto} and recently Score-based Generative Models (SGMs), which aim to generate high-quality samples, sample quickly, and have adequate mode coverage \citep{xiao2021tackling}. 
Over recent years, SGMs have become the \textit{de facto} method for image generation due to the quality and diversity of generated images \citep{dhariwal2021diffusion}. 
Generative models can be integrated into the reconstruction process as data-driven priors\blue{, through the learnt image manifold, but which are independent of the specifics of the forward model, cf. \citep{dimakis_2022}.
For example, a generative model trained on PET brain data would not be appropriate for MRI reconstruction of knees, but would be appropriate for the reconstruction of other PET brain images. In the latter case, by suitably changing the forward model, the generative model could be used across scanners and noise levels.} 

SGMs have been applied to CT and MR image reconstruction \citep{song2021solving}. These reconstructions condition the SGM image generation on measurements, and balance the consistency with measurements versus consistency with the SGM learnt image manifold \citep{kobler2023learning}. 
There are different methods to enforce measurement consistency of the reconstructions, which can be broadly classified into gradient based methods \citep{jalal2021robust, chung2023noise} and projection based methods \citep{song2021solving, chung2022score, chung2023fast}. 
Recently, denoising diffusion models (discrete variants of SGMs) were used for PET image denoising \citep{gong2022pet}. Instead, our work focuses on PET image reconstruction, and we present the following contributions: 
\begin{itemize}
    \item We develop a novel algorithmic framework building upon SGMs that carefully addresses the challenges inherent to PET. To do so, we modify the conditional sampling method \citep{chung2023fast, zhu2023denoising}, recently proposed for inverse problems with Gaussian noise, for PET image reconstruction. This is achieved with a penalised Maximum A Posteriori (MAP) estimator computed with an accelerated algorithm that evaluates subsets of the measurements.   
    \item We leverage additional MR images to enhance the proposed framework, leading to improved image quality that better agrees with the measured data.
    \item We scale the approach to 3D PET reconstruction.
\end{itemize}

The proposed method is tested on multiple noise levels, radio-tracers, and in both 2D and 3D settings with an SGM trained on patient-realistic BrainWeb data without lesions \citep{Collins:1998brainweb}. 
In addition to data without lesions, we test on out-of-distribution (OOD) data with lesions to validate method robustness.
The rest of the paper is structured as follows. In Section~\ref{sec:background} we provide the background on PET reconstruction and SGMs. In particular, we present different methods for using SGMs in image reconstruction. In Section~\ref{sec:appl_to_PET} we propose modifications needed to apply SGMs for PET reconstruction. We describe the experimental setting in Section \ref{sec:experimental_setup}, and present and discuss the results in Section~\ref{sec:results}. 
\blue{The code is publicly available on Github~\footnote{\url{https://github.com/Imraj-Singh/Score-Based-Generative-Models-for-PET-Image-Reconstruction}}, and the dataset on Zenodo\footnote{\url{https://zenodo.org/records/10509379}}}. 

\section{Background}
\label{sec:background}

\subsection{Fundamentals of Positron Emission Tomography Reconstruction}
\label{sec:fundamentals_of_pet}
PET measurements are the result of a low-count photon-counting process. The true forward process, from tracer-distribution to photon detection, is approximated by the forward model defined in Eq.~\eqref{eq:pet_inverse_problem}.
The likelihood 
of the measured photon counts, for an unknown tracer distribution, can be modelled by an independent Poisson distribution. One of the first methods developed for estimating the tracer distribution through a Poisson model was maximum likelihood. This selects an image~\mbox{$\Bx \in\mathbb{R}_{\geq 0}^n$} by maximising the Poisson Log-Likelihood (PLL) function, given by
\begin{equation}
\label{eq:nll_pet}
    L(\By|\Bx) = \sum_{i=1}^m y_i \log([\BA\Bx+\Bbarb]_i) - [\BA\Bx+\Bbarb]_i-\log(y_i!).
\end{equation}
By maximising the PLL, the Maximum Likelihood Estimate (MLE) is obtained. A particularly important algorithm for computing the MLE is Expectation Maximisation (EM) \citep{shepp1982maximum}. However, due to its slow convergence, acceleration is sought through splitting the PLL into a sum of $n_{\mathrm{sub}}\ge1$ sub-objectives. This gives rise to the greatly sped-up Ordered Subset Expectation Maximisation (OSEM) \citep{HudsonLarkin:1994} algorithm.
Because of the ill-conditioning of PET reconstruction, the MLE tends to overfit to measurement noise. To address the ill-conditioning and improve reconstruction quality, it is common practice to regularise the reconstruction problem via the use of an image-based prior. This gives rise to the MAP objective
\begin{equation} \label{eq:map_obj}
    \Phi(\mathbf{x}) = L(\mathbf{y}|\mathbf{x}) + \lambda R(\mathbf{x}),
\end{equation}
where $R(\Bx)$ is the log of a chosen image-based prior with penalty strength~$\lambda$. Block-Sequential Regularised Expectation Maximisation (BSREM) \citep{DePierro2001BSREM, ahn2003BSREM} is an iterative algorithm, globally convergent under mild assumptions, that applies the subset idea of OSEM to the MAP objective. 
For $\Phi(\Bx)=\sum_{j=1}^{n_{\mathrm{sub}}} \Phi_j(\Bx)$, where $\Phi_j$ is the sub-objective~$\Phi_j(\Bx) = L_j(\By|\Bx) + \lambda R(\Bx)/n_\mathrm{sub}$, and $L_j$ is the likelihood for a subset of the measurements. The BSREM update iterations are given by
\begin{equation}
\label{eq:bsrem}
\mathbf{x}^{i+1} = P_{\mathbf{x}\geq 0} \left[ \mathbf{x}^{i} + \alpha_{i} \mathbf{D}(\mathbf{x}^{i}) \nabla \Phi_{j} (\mathbf{x}^i) \right]\quad i\ge0,
\end{equation}
where $P_{\mathbf{x}\geq 0} [\cdot]$ denotes the non-negativity projection, $i$ is the iteration number, and index $j = (i \mod n_{\mathrm{sub}})+1$ cyclically accesses sub-objectives. The preconditioner is $\mathbf{D}(\Bx^{i}) = \mathrm{diag} \left( {\max(\Bx^{i}, \delta)}/{\mathbf{A}^\top \mathbf{1}} \right)$, where $\mathbf{1}\in\mathbb{R}^m$ denotes the vector with all entries equal to 1,  $\delta$ is a small positive constant to ensure positive definiteness, and $\mathbf{A}^\top$ the matrix transpose.
The quantity $\mathbf{A}^\top \mathbf{1}$ is referred to as the sensitivity image. The step-sizes are $\alpha_{i} = \alpha_0/ (\zeta \lfloor i/ n_{\mathrm{sub}}\rfloor+ 1)$, where $\alpha_0=1$ and $\zeta$ is a relaxation coefficient. A common regulariser for PET reconstruction is the Relative Difference Prior (RDP) \citep{nuyts2002concave}, see Appendix \ref{app:baseline_methods} for details. The gradient of the RDP is scale-invariant as it is computed using the ratio of voxel values. This partially overcomes the issue with the wide dynamic range observed in emission tomography images, helping to simplify the choice of the penalty strength across noise levels.

\subsection{Score-based Generative Models} \label{sec:sgm}

SGMs have emerged as a state-of-the-art method for modelling, and sampling from, high-dimensional image distributions \citep{song2020score}. They reinterpret denoising diffusion probabilistic modelling \citep{sohl2015deep, ho2020denoising} and score-matching Langevin dynamics \citep{song2019generative} through the lens of Stochastic Differential Equations (SDE). SGMs are often formulated by prescribing a forward diffusion process defined by an It\^o SDE
\begin{equation}
\label{eq:forward_sde}
    d \Bxt = \Bf(\Bxt, t) d t + g(t) d \mathbf{w}_t, \quad \Bx_0 \sim p_0 := \pprior, 
\end{equation}
where $\{\Bx_t\}_{t \in [0,T]}$ is a stochastic process indexed by time $t$ and $\pprior$ is the image distribution.
Each random vector $\Bx_t$ has an associated time-dependent density $p(\Bx_t)$. To emphasise that the density is a function of $t$ we write $p_t(\Bx_t) := p(\Bx_t)$. The multivariate Wiener process $\{\Bw_t\}_{t\ge0}$ is the standard Brownian motion. Starting at the image distribution $\pprior$, the drift function $\Bf(\cdot, t): \R^n \to \R^n$ and the diffusion function $g:\R^n \to \R$ are chosen such that the terminal distribution at $t=T$ approximates the standard Gaussian, $p_T \approx \mathcal{N}(0,I)$. 
Thus, the forward diffusion process maps the image distribution $\pprior$ to a simple, tractable distribution. The aim of SGMs is to invert this process, i.e. start at the Gaussian distribution and go back to the image distribution $\pprior$. Under certain conditions on $\Bf$ and $g$, a reverse diffusion process can be defined \citep{anderson1982reverse}
\begin{equation}\label{eq:reverse_sde}
    d \Bxt = [\Bf(\Bxt, t) - g(t)^2 \nabla_\Bx \log \ppriort(\Bxt)]  d t + g(t) d \Bar{\mathbf{w}}_t,
\end{equation}
that runs backwards in time. The Wiener process $\{ \Bar{\mathbf{w}}_t\}_{t \ge 0}$ is time-reversed Brownian motion, and the term $\nabla_\mathbf{x} \log \ppriort(\Bxt)$ is the score function.
Denoising Score Matching (DSM) \citep{vincent2011connection} provides a methodology for estimating $\nabla_{\Bx} \log \ppriort(\Bxt)$ by matching the transition densities $p_t(\Bxt|\Bxzero)$ with a time-conditional neural network $s_\theta(\Bxt, t)$, called the score model, parametrised by $\theta$. The resulting optimisation problem is given by
\begin{equation}
\label{eq:denoising_score_matching}
    \min_\theta \left\{ L_\text{DSM}(\theta) = \mathbb{E}_{t \sim U[0,T]} \mathbb{E}_{\Bxzero \sim \pprior} \mathbb{E}_{\Bxt \sim p_t(\Bxt|\Bxzero)} \left[ \omega_t \| s_\theta(\Bxt, t) - \nabla_{\Bx} \log p_t(\Bxt|\Bxzero) \|_2^2 \right] \right\},
\end{equation}
 where $\omega_t > 0$ are weighting factors, balancing the scores at different time steps. For general SDEs, the loss $ L_\text{DSM}(\theta)$ may still be intractable, since it requires access to the transition density $p_t(\Bxt|\Bxzero)$. However, for SDEs with an affine linear drift function, $p_t(\Bxt|\Bxzero)$ is a Gaussian and thus can be given in closed form \citep{sarkka2019applied}. Throughout the paper, we use $T=1$ and the variance preserving SDE \citep{ho2020denoising} given by 
\begin{equation}
    \label{eq:vp-sde}
    d \Bxt = - \frac{\beta(t)}{2} \Bxt d t + \sqrt{\beta(t)} d \mathbf{w}_t,
\end{equation}
where $\beta(t): [0,1] \to \R_{>0}$ is an increasing function defining the noise schedule. We use $\beta(t) = \betamin + t (\betamax - \betamin)$ giving the transition kernel $p_t(\Bxt|\Bxzero)= \mathcal{N}(\Bxt; \gamma_t \Bxzero, \nu_t^2 I)$ with coefficients $\gamma_t, \nu_t \in \R$ computed from  drift and diffusion coefficients,
see Appendix \ref{app:training_SGM} for details. 
Generating samples with the score model $s_\theta(\Bxt,t)$ as a surrogate requires solving the reverse SDE \eqref{eq:reverse_sde}, with the score model $s_\theta(\Bxt,t)$ in place of $\nabla_\Bx \log \ppriort(\Bxt)$
\begin{equation}
\label{eq:reverse_sde_with_model}
    d \Bxt = [\Bf(\Bxt, t) - g(t)^2 s_\theta(\Bxt, t)]  d t + g(t) d \Bar{\mathbf{w}}_t.
\end{equation}
Drawing samples from the resulting generative model thus involves two steps. First, drawing a sample from the terminal distribution $\Bx_1 \sim \mathcal{N}(0,I) \approx p_1$, and second, initialising the reverse SDE~\eqref{eq:reverse_sde_with_model} with~$\Bx_1$ and simulating backwards in time until $t=0$. The latter can be achieved by Euler-Maruyama schemes or predictor-corrector methods \citep{song2020score}.

\subsubsection{Denoising diffusion implicit models}
Simulating the reverse SDE can be computationally expensive as a fine time grid is often necessary to produce realistic samples. Denoising Diffusion Implicit Models (DDIMs) \citep{song2020denoising} were introduced to allow faster sampling, and build upon a result by Tweedie \citep{efron2011tweedie} to approximate the expectation $\E[\Bxzero | \Bxt]$ via the score model $s_\theta(\Bxt, t)$ as 
\begin{equation}
    \label{eq:tweedie_denoising}
     \E[\Bxzero | \Bxt] = \frac{\Bxt + \nu_t^2 \nabla_\Bx \log \ppriort(\Bxt)}{\gamma_t} \approx  \frac{\Bxt + \nu_t^2 s_\theta(\Bxt, t)}{\gamma_t} := \Tweedie(\Bxt),
\end{equation}
\blue{where the positive scalars $\gamma_t$ and $\nu_t^2$ are the coefficients for the mean and covariance, respectively, defining the transition kernel, cf. \eqref{eqn:gammanu_t} in the Appendix for details.}
DDIM defines a non-Markovian sampling rule, which uses both the current sample $\Bxt$ and Tweedie's estimate $\Tweedie(\Bxt)$ to create an accelerated sampler. Let $0 = t_{k_1} \le \dots \le t_{k_N} = 1$ be the time discretisation. The DDIM sampling update can be written as
\begin{equation}
    \label{eq:ddim_sampling}
    \begin{split}
         \Bx_{t_{k-1}} &=  \gamma_{t_{k-1}} \Tweedie(\Bx_{t_k}) + \text{Noise}(\Bx_{t_k}, s_\theta) + \eta_{t_k} \Bz, \quad \Bz \sim \mathcal{N}(0,I) \\ 
          &\text{ with }\text{Noise}(\Bx_{t_k}, s_\theta) := - \nu_{t_k} \sqrt{ \nu_{t_{k-1}}^2 - \eta_{t_k}^2 } s_\theta(\Bx_{t_k}, t_k).
    \end{split}
\end{equation}
The sampling rule can be split into a denoising step (predicting $\Tweedie(\Bx_{t_k})$ using the score model), and adding an appropriate amount of noise back. Thus, the sampling mimics an iterative refinement process, as the prediction of the denoised estimate $\Tweedie(\Bx_{t_k})$ will be more accurate for smaller~$t_k$.  Different choices of $\eta_t$ result in different sampling schemes. We choose $\eta_{t_k} = \eta \beta_{t_k} $ with a hyperparameter~\mbox{$\eta \in [0,1]$}, controlling the amount of stochasticity in the sampling, and $\beta_{t_k}=\nu_{t_{k-1}}/\nu_{t_{k}} \sqrt{1 -\gamma_{t_k} / \gamma_{t_{k-1}}}$
\citep{song2020denoising}.

\subsection{Using Score-based Generative Models for Inverse Problems}
\label{sec:SGMforIP}
The goal of the Bayesian framework of inverse problems is to estimate the posterior $\pPost(\Bx|\By)$, i.e. the conditional distribution of images $\Bx$ given noisy measurements $\By$. Using Bayes' theorem the posterior can be factored into
\begin{equation}
    \label{eq:bayes}
    \pPost(\Bx|\By) \propto \pLkhd(\By | \Bx) \pprior(\Bx), \quad \nabla_{\Bx} \log \pPost(\Bx|\By) = \nabla_{\Bx} \log \pLkhd(\By | \Bx) + \nabla_{\Bx} \log \pprior(\Bx),
\end{equation}
where $\pLkhd$ denotes the likelihood and $\pprior$ the prior given by the image distribution. We can set up a generative model for the posterior in the same way as for the prior $\pprior$ in Section~\ref{sec:sgm} by defining a forward SDE which maps the posterior to random noise. To generate a sample from the posterior~$\pPost(\Bx|\By)$, we can simulate the corresponding reverse SDE
\begin{equation}
\label{eq:cond_reverse_sde_posterior}
    d \Bxt = [\Bf(\Bxt, t) - g(t)^2  \nabla_{\Bx} \log p_t(\Bxt|\By)]  d t + g(t) d \Bar{\mathbf{w}}_t,
\end{equation}
where we need access to the time-dependent posterior $\nabla_{\Bx} \log p_t(\Bxt|\By)$. Similar to Eq.~\eqref{eq:bayes}, we use Bayes' theorem for the score of the posterior and decompose $\nabla_{\Bx} \log p_t(\Bxt|\By)$ into a prior and a likelihood term, where the former is approximated with the trained score model
\begin{equation}
    \begin{split}
        \nabla_{\Bx} \log p_t(\Bxt|\By) &= \nabla_{\Bx} \log \ppriort(\Bxt) + \nabla_{\Bx} \log \pLkhdt(\By|\Bxt) \\ 
        &\approx s_\theta(\Bxt, t) + \nabla_{\Bx} \log p_t(\By|\Bxt).
    \end{split}
\end{equation}
Substituting the above approximation into~\eqref{eq:cond_reverse_sde_posterior}, we obtain
\begin{equation}
\label{eq:cond_reverse_sde_with_model}
    d \Bxt = [\Bf(\Bxt, t) - g(t)^2 ( s_\theta(\Bxt, t) + \nabla_{\Bxt} \log \pLkhdt(\By|\Bxt))]  d t + g(t) d \Bar{\mathbf{w}}_t.
\end{equation}
We can recover approximate samples from the posterior $\pPost(\Bx|\By)$, by simulating the reverse SDE~\eqref{eq:cond_reverse_sde_with_model}. Through iterative simulation of the reverse SDE with varying noise initialisations, we can estimate moments of the posterior distribution. As is common practice in the field \citep{song2021solving, chung2022score,jalal2021robust} we use one sample for the reconstruction, due to computational costs of repeatedly solving the reverse SDE. In addition to the score model $s_\theta(\Bxt, t)$, we need the score of the time-dependent likelihood $\nabla_{\Bx} \log \pLkhdt(\By|\Bxt)$. At the start of the forward SDE (for $t=0$), it is equal to the true likelihood $\pLkhd$. However, for $t>0$ the score $\nabla_{\Bx} \log \pLkhdt(\By|\Bxt)$ is intractable to compute exactly and different approximations have been proposed. In  \citep{jalal2021robust, ramzi2020denoising}, this term was approximated with the likelihood $\pLkhd$ evaluated at the noisy sample $\Bxt$ with time-dependent penalty strength $\lambda_t$
\begin{equation}
    \label{eq:naive_approx}
    \nabla_{\Bx} \log \pLkhdt(\By | \Bxt) \approx \lambda_t \nabla_{\Bx} \log \pLkhd(\By |\Bxt). 
\end{equation}
We refer to Eq.~\eqref{eq:naive_approx} as the \Naive~ approximation. The Diffusion Posterior Sampling (DPS) \citep{chung2023noise} uses Tweedie's formula to obtain $\hat{\Bx}_0(\Bxt) \approx \E[\Bxzero | \Bxt]$ and approximates $\nabla_{\Bx} \log \pLkhdt(\By | \Bxt)$ by
\begin{equation}
    \label{eq:dps_approx}
    \nabla_{\Bx} \log \pLkhdt(\By | \Bxt) \approx \nabla_{\Bx} \log \pLkhd(\By |\Tweedie(\Bxt)), 
\end{equation}
where $\nabla_{\Bx}$ denotes taking derivative in $\Bx_t$ (instead of $\hat {\Bx}_0$). It was shown that this approximation leads to improved performance for several image reconstruction tasks \citep{chung2023noise}. However, DPS comes with a higher computational cost, due to the need to back-propagate the gradient through the score model. 

Recently, several works proposed modifying the DDIM sampling rule in Eq.~\eqref{eq:ddim_sampling} for conditional generation \citep{zhu2023denoising, chung2023fast}. These methods generally consist of three steps:~(1) estimating  the denoised image $\Bxzero$ using Tweedie's estimate $\Tweedie(\Bx_{t_k})$; (2) updating $\Tweedie(\Bx_{t_k})$ with a data consistency step using the measurements $\By$; and (3) adding the noise back, according to the DDIM update rule, in order to get a sample for the next time step $t_{k-1}$. Importantly, with this approach there is no need to estimate the gradient of the time-dependent likelihood $\nabla_{\Bx} \log \pLkhdt(\By | \Bxt)$ as data consistency is only enforced on Tweedie's estimate at $t=0$. 
These conditional DDIM samplers differ most greatly in the implementation of the data consistency update. Decomposed Diffusion Sampling (DDS)  \citep{chung2023fast} proposes to align Tweedie's estimate with the measurements by running $p$ steps of a Conjugate Gradient (CG) scheme for minimising the negative log-likelihood at each sampling step. Let $\text{CG}^{(p)}(\Tweedie)$ denote the $p$-th CG update initialised with $\Tweedie(\Bx_{t_k})$. This can be seen as an approximation to the conditional expectation, i.e. $\E[\Bxzero|\Bxt, \By] \approx \text{CG}^{(p)}(\Tweedie)$ \citep{ravula2023optimizing}. Using this approximation, the update step for DDS can be written as 
\begin{equation}
\label{eq:dds_update}
    \Bx_{t_{k-1}} = \gamma_{t_{k-1}} \text{CG}^{(p)}(\Tweedie)  + \text{Noise}(\Bx_{t_k}, s_\theta) + \eta_{t_k} \Bz, \text{ with } \Bz \sim \mathcal{N}(0, I), 
\end{equation}
where the introduction of the conditional expectation offers us the possibility to explore different approximations specific for PET image reconstruction.

\section{PET-specific Adaptations for SGMs} \label{sec:appl_to_PET}

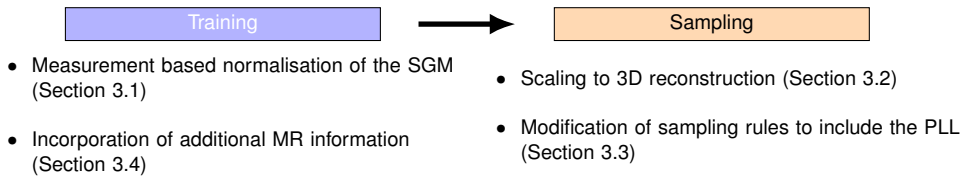
\begin{figure}[h!]
\centering
\scriptsize
\begin{tikzpicture}[node distance=6.5cm, font=\sffamily]
    \node[draw, rectangle, fill=blue!30, text=white, text width=4cm, align=center] (training) {Training};
    \node[below of=training, node distance=1cm, text width=7cm, align=left] (training-desc) {
    \begin{itemize}  
        \item Measurement based normalisation of the SGM (Section \ref{sec:instance_norm})
        \item Incorporation of additional MR information (Section \ref{sec:guided_reco})
    \end{itemize}
    };

    \node[draw, rectangle, fill=orange!30, text width=4cm, align=center, right of=training] (sampling) {Sampling};
    \node[below of=sampling, node distance=1cm, text width=7cm, align=left] (sampling-desc) {
    \begin{itemize}
        \item Scaling to 3D reconstruction (Section \ref{sec:scaling_to_3d})
        \item Modification of sampling rules to include the PLL (Section \ref{sec:conditional_sampling})
    \end{itemize}
    };
    \draw[->, -{Latex[length=10pt,width=8pt]}, line width=2pt]  ([xshift=0.5cm]training.east) -- ([xshift=-0.5cm]sampling.west);
\end{tikzpicture}
\caption{\blue{\vspace{-1em}Schematic illustration of the modification for training and sampling steps of SGMs.}}
\vspace{-2em}
\label{fig:changes_to_sampling}
\end{figure}

To apply SGMs to PET reconstruction, several key components of the pipeline in Section \ref{sec:sgm} have to be modified in order to incorporate PET-specific constraints. Namely, we introduce measurement-based normalisation of the input to the score model, and explain how to apply a score model trained on 2D slices for 3D reconstruction. Additionally, we adapt the sampling methods from Section \ref{sec:SGMforIP} to incorporate the Poisson noise model. Finally, we demonstrate that the SGM framework allows for the incorporation of additional information, e.g. MR images, by using classifier-free guidance \citep{ho2022classifier}. \blue{The overall adaption steps are summarised in Fig. \ref{fig:changes_to_sampling}.}
\subsection{Measurement-based Normalisation}
\label{sec:instance_norm}
The intensity of the unknown tracer distribution in emission tomography can significantly vary across different scans, resulting in a high dynamic range that poses challenges for deep learning approaches. Neural networks may exhibit bias toward intensity levels that appear more frequently in the training set. Consequently, the network might struggle to handle new images with unseen intensity levels, leading to instability in the learning and evaluation process \citep{tran2020practical}. For SGMs the intensity range of images must be predefined to ensure that the forward diffusion process converges to a standard Gaussian distribution and to stabilise the sampling process \citep{lou2023reflected}. Input normalisation is a standard deep learning methodology to deal with intensity shifts and normalise the inputs to the network. In a similar vein, we propose a PET-specific normalisation method to ensure that the score model $s_\theta(\Bxt, t)$ is able to estimate the score function of images with arbitrary intensity values. Namely, we normalise each training image $\Bxzero$ to ensure voxel intensities are within a certain range. To do this 
we introduce a training normalisation factor $c_{\text{train}}$ that when applied ensures the average emission per emission voxels (a voxel with non-zero intensity value) is~1. This is computed as
\begin{equation}\label{eqn:c_train} 
    c_{\text{train}} = c(\Bxzero) :=\frac{\sum_{j=1}^m [\Bxzero]_j}{\#\{j: [\Bxzero]_j>0\}},
\end{equation}
where the numerator captures the total emission in the image, and the denominator is the number of emission voxels.
The normalisation factor is incorporated into the DSM training objective function by rescaling the initial image, yielding the objective
\begin{equation}
\label{eq:denoising_score_matching_scale_robust}
\mathbb{E}_{t \sim U[0,T]} \mathbb{E}_{\Bxzero \sim \pprior} \mathbb{E}_{\Bz \sim \mathcal{N}(0,I)} \mathbb{E}_{c \sim U[\frac{c_{\rm train}}{2}, \frac{3c_{\rm train}}{2}]} \left[ \omega_t \left\| s_\theta \left(\tilde \Bx_t, t \right) - \nabla_{\Bx} \log p_t(\tilde \Bx_t|\Bxzero/c )\right\|_2^2 \right],
\end{equation}
with $\tilde \Bx_t = \gamma_t \Bxzero/c + \nu_t \Bz$. Compared with Eq.~\eqref{eq:denoising_score_matching}, the scale-factor in range $c \sim U[\frac{c_{\rm train}}{2}, \frac{3c_{\rm train}}{2}]$ is used to encourage the score model to be more robust with respect to misestimations of the normalisation constant during sampling.

An analogue of Eq.~\eqref{eqn:c_train} is unavailable during the sampling, and thus a surrogate is required. This is obtained through an approximate reconstruction, computed using a single epoch of OSEM from a constant non-negative initialisation. 
The resulting sampling normalisation factor is given by
\begin{equation}
    c_{\text{OSEM}} = \frac{\sum_{j=1}^m [\Bx_{\text{OSEM}}]_j}{\#\{j: [\Bx_{\rm OSEM}]_j>Q_{0.01}\}},
\end{equation}
where $Q_{0.01}$ defines the $1\%$ percentile of $\Bx_{\text{OSEM}}$ values. This threshold is heuristically chosen to ensure that noise and reconstruction artefacts do not cause an over-estimation of the number of emission voxels. 
In the reconstruction process the normalisation constant $c_{\rm OSEM}$ is applied as a factor scaling the time-dependent likelihood, giving 
\begin{equation}
\label{eq:cond_reverse_sde_with_model_and_norm}
    d \Bxt = [\Bf(\Bxt, t) - g(t)^2 ( s_\theta(\Bxt, t) +  \nabla_{\Bx} \log \pLkhdt(\By|c_{\text{OSEM}}\Bxt))]  d t + g(t) d \Bar{\mathbf{w}}_t.
\end{equation}
At final time step $t=0$, the output $\Bx$ is rescaled by $c_{\rm OSEM}$ to recover the correct intensity level.

\subsection{Scaling to 3D Reconstruction}
\label{sec:scaling_to_3d}

While some SGM studies deal with fully 3D image generation \citep{pinaya2022brain}, the majority of work is restricted to 2D images. This is largely due to the fact that the learning of full 3D volume distributions is computationally expensive and requires access to many training volumes. Therefore, we propose to train the score model on 2D axial slices and use a specific decomposition of the conditional sampling rules to apply the model for 3D reconstruction. Upon simulating the conditional reverse SDE in Eq.~\eqref{eq:cond_reverse_sde_with_model} using the Euler-Maruyama approach, we arrive at the iteration rule
\begin{align}
\label{eq:em_decomposed_1}
    &\tilde \Bx_{t_{k-1}} = \Bx_{t_k} + \big[\Bf(\Bx_{t_k}, t_k) - g(t_k)^2 s_\theta(\Bx_{t_k}, t_k) \big] \Delta t  + g(t_k) \sqrt{|\Delta t|} \Bz, \quad \Bz \sim \mathcal{N}(0, I), \\ 
    &\Bx_{t_{k-1}} = \tilde \Bx_{t_{k-1}}  - g(t_k)^2  \nabla_{\Bx
    } \log p_{t_k}(\By | \Bxt_k) \Delta t,\label{eq:em_decomposed_2}
\end{align}
using an equidistant time discretisation $0 = t_{k_1} \le \dots \le t_{k_N} = 1$ for $N \in \N$, with a time step $\Delta t = - 1/N$. We split the Euler-Maruyama update into two equations to highlight the influences of the score model and the measurements $\By$. First, Eq.~\eqref{eq:em_decomposed_1} is the Euler-Maruyama discretisation for the unconditional reverse SDE, see Eq.~\eqref{eq:reverse_sde_with_model}. This update is independent of the measurements $\By$ and can be interpreted as a prior update, increasing the likelihood of $\Bx_{t_k}$ under the SGM. The second step in Eq.~\eqref{eq:em_decomposed_2} is a data consistency update, pushing the current iterate to better fit the measurements. Notably, this step is fully independent of the score model. This strategy was developed for 3D reconstruction, focusing on sparse view CT and MRI \citep{chung2022solving}. It was proposed to apply the prior update in Eq.~\eqref{eq:em_decomposed_1} to all slices in the volume independently and use the 3D forward model in the data consistency step. Further, a regulariser in the direction orthogonal to the slice was introduced, to improve consistency of neighbouring slices. However, applying this approach to the Euler-Maruyama discretisation results in slow sampling times as a small time step $|\Delta t|$ is necessary. 
To accelerate the sampling of high quality samples, we propose to use the DDS update in Eq.~\eqref{eq:dds_update} that uses a similar decomposition of independent score model updates to axial slices, and 3D data consistency updates. Additionally, we accelerate data consistency updates by splitting the measurement data into ordered subsets and applying the forward model block-sequentially. The details are explained below. 

\subsection{Modifications of Sampling Methods}
\label{sec:conditional_sampling}

The sampling schemes and approximations in Section~\ref{sec:SGMforIP} were originally proposed for inverse problems with Gaussian noise. The work on DPS \citep{chung2023noise} also considers inverse problems with Poisson noise, but utilises a Gaussian approximation to the Poisson noise, which is known to be unsuitable for PET reconstruction in the event of the low photon count \blue{\citep{BerteroBoccacci:2009,HohageWerner:2016}}. To apply the \Naive{} approximation or the DPS approach to Poisson inverse problems, one could simply replace the Gaussian log-likelihood with the PLL in Eq.~\eqref{eq:naive_approx} and Eq.~\eqref{eq:dps_approx}. However, PLL and its gradient are only defined for non-negative values. Therefore, we have to introduce a non-negativity projection into the sampling to ensure that the gradient of the PLL can be evaluated. In the context of guided diffusion, it was proposed to project the iterates $\Bx_{t_k}$ to a specified domain after each sampling step \citep{li2022diffusion, saharia2022photorealistic}. In our case this would require thresholding all negative values. However, this creates a mismatch between the forward and reverse SDEs. It was observed that this mismatch results in artefacts in the reconstructions and may even lead to divergence of the sampling \citep{lou2023reflected}. \blue{In our experiments, thresholding all negative values of $\Bx_{t_k}$ leads to a divergence of the sampling process.} Therefore, we propose to only threshold the input to the PLL, i.e.  with $L$ being the PLL, see Eq. \eqref{eq:nll_pet}, for the \PETNaive{} approximation we use 
\begin{equation}
    \label{eq:naive_pet}
    \nabla_{\Bx} \log \pLkhdt(\By | \Bxt) \approx \lambda_{t}^\text{\Naive{}} \nabla_{\Bx} L(\By | c_{\text{OSEM}} P_{\Bx \geq 0}[\Bx_{t}]),
\end{equation}
and likewise for \PETDPS
\begin{equation}
    \label{eq:dps_pet}
    \nabla_{\Bx} \log \pLkhdt(\By | \Bxt) \approx \lambda_{t}^\text{DPS} \nabla_{\Bx} L(\By | c_{\text{OSEM}} P_{\Bx \geq 0}[\Tweedie(\Bx_{t})]).
\end{equation}
Note that this leads to a perturbed likelihood gradient that is not computed on the true iterate $\Bx_{t}$, but only on the projection. In order to reconstruct the PET image we have to solve the reverse SDE using the specific approximation (\PETNaive~ or \PETDPS) as the likelihood term. This usually requires around $1000$ sampling steps to produce an appropriate reconstruction and results in impractically long reconstruction times for 3D volumes.

To reduce the reconstruction times we propose to modify the conditional DDIM sampling rule, which we call \PETDDS, similar to the DDS framework \citep{zhu2023denoising, chung2023fast}, cf. Section~\ref{sec:conditional_sampling}. This circumvents the usage of $\nabla_\Bx \log \pLkhdt(\By | \Bxt)$, instead enforcing data consistency for Tweedie's estimate $\Tweedie(\Bx_{t_k})$. For PET reconstruction we propose to implement this data consistency with a MAP objective, leading to the \PETDDS{} update
\begin{align}
    \Bx^0_{t_k} &= \Tweedie(\Bx_{t_k}) \label{eq:pet_dds_tweedie} \\ 
    \Bx^{i+1}_{t_k} &= P_{\mathbf{x}\geq 0} \left[  \Bx^{i}_{t_k} + \mathbf{D}(\Bx^{i}_{t_k})  \nabla_\Bx\Phi_{j}(\Bx^i_{t_k})\right]\label{eq:pet_dds_map}  \\ 
    &\quad \quad i = 0, \dots, p-1 \nonumber\\ 
    \Bx_{t_{k-1}} &= \gamma_{t_{k-1}} \Bx^{p}_{t_k} + \text{Noise}(\Bx_{t_k}, s_\theta) + \eta_{t_k} \Bz, \quad \Bz \sim \mathcal{N}(0, I), \label{eq:pet_dds_ddim}
\end{align}
where \blue{$\mathbf{D}(\Bx) = \mathrm{diag} \left\{ {\max(\Bx, 10^{-4})}/{\mathbf{A}^\top \mathbf{1}} \right\}$ is the preconditioner and } the sub-objective is 
\begin{equation} \label{eq:pet_dds_objective}
     \Phi_{j}(\Bx^i) =  L_{j} (\By | c_\text{OSEM} \Bx^{i}) + (\lambda^\text{RDP}  R_z(\Bx^i)- \lambda^\text{DDS}\| \Bx^i - \Tweedie \|_2^2)/n_{\mathrm{sub}}.
\end{equation}
The sub-objective index $j=j(i)$ is given by $ j = (p(N-k)+i \mod n_{\rm{sub}}) + 1$, which cyclically accesses sub-objectives. The RDP used for 3D data $R_z$ is applied in the $z$-direction, perpendicular to the axial slice, see Appendix \ref{app:baseline_methods} for more details. The prior in Eq.~\eqref{eq:pet_dds_objective} consists of two components: one anchoring to the Tweedie's estimate $\| \Bx - \Tweedie \|_2^2$, and the other RDP in the $z$-direction $R_z(\Bx)$. The components have associated penalty strengths $\lambda^\text{RDP}$ and $\lambda^\text{DDS}$, respectively.

In a \PETDDS{} update we first independently compute Tweedie's estimate based on $\Bx_{t_k}$ for each axial slice (Eq.~\ref{eq:pet_dds_tweedie}). Tweedie's estimate $\Tweedie$ impacts the reconstruction in two ways: first through the Tikhonov regulariser scaled with $\lambda^\text{DDS}$, and second as the initial value for the projected gradient descent in Eq.~\eqref{eq:pet_dds_map}. Through running $p$ steps of projected gradient descent consistency is balanced between a PLL on measurements, RDP in the $z$-direction, and Tweedie's estimate (Eq.~\ref{eq:pet_dds_objective}). To speed up computation of the objective gradient, the objective is split into sub-objectives and the gradient of the log-likelihood is evaluated using only subsets of the measurements $\By$, similar to the BSREM update in Eq.~\eqref{eq:bsrem}. The subsets are partitioned in a staggered configuration and are ordered with a Herman-Meyer order \citep{Herman1993}. Eq.~\eqref{eq:pet_dds_ddim} is the DDIM update applied to the conditioned Tweedie estimate $\Bx^{p}$, where the score update is again applied independently for each axial slice, here the notation of $\text{Noise}(\Bx_{t_k}, s_\theta)$ is overloaded. The DDIM update gives $\Bx_{t_{k-1}}$, and these \PETDDS{} updates repeat until $t_{0}=0$ giving reconstruction $\hat{\Bx}$.

\begin{table}[h!]
\centering
\vspace{-1em}
\caption{\blue{Summary of different sampling schemes proposed for PET. 
} \vspace{-1em}}
\begin{tabular}{l|c|c|c}
          \textbf{Method} & \textbf{Sampling type} & \textbf{Data consistency with $L$ in \eqref{eq:nll_pet}} & \textbf{Algorithm}  \\ \midrule
\textbf{\PETNaive} & Euler-Maruyama \eqref{eq:em_decomposed_1} & $L(\By | c_{\text{OSEM}} P_{\Bx \geq 0}[\Bx])$ \eqref{eq:naive_pet} &  Algo. \ref{alg:nav}         \\
\textbf{\PETDPS}   & Euler-Maruyama \eqref{eq:em_decomposed_1} & $L(\By | c_{\text{OSEM}} P_{\Bx \geq 0}[\Tweedie(\Bx)])$ \eqref{eq:dps_pet} &  Algo. \ref{alg:dps}\\
\textbf{\PETDDS}   & DDIM \eqref{eq:ddim_sampling}         & $L_{j} (\By | c_\text{OSEM} \Bx)$ \eqref{eq:pet_dds_map} and \eqref{eq:pet_dds_objective} & Algo. \ref{alg:dds}\\
\bottomrule
\end{tabular}
\label{tab:petmethods}
\end{table}

\blue{In Table \ref{tab:petmethods}, we list the details and differences of the proposed score-based schemes for PET.}

\subsection{MR Image Guided Reconstruction}
\label{sec:guided_reco}
In recent years, several regularisation methods have been proposed which leverage the availability of additional MR images to improve PET image reconstruction \citep{ehrhardt2016pet, bai2013magnetic, somayajula2010pet}. These studies often encode anatomical features of the MR image as edges or level sets and build hand-crafted regularisers based on these encoded features. This is commonly referred to as guided reconstruction \citep{Ehrhardt2021}, where the MR image is first reconstructed and is then used in the PET reconstruction pipeline. The SGM approach can be modified for guided reconstruction. In this setting we can use Bayes' theorem to express the posterior 
\begin{equation}
        \nabla_\Bx \log \pPost(\Bx|\By, \BxMR) = \nabla_\Bx \log \pLkhd(\By|\Bx) + \nabla_\Bx \log \pprior(\Bx | \BxMR) ,
\end{equation}
assuming that both $\By$ and $\BxMR$ are conditionally independent given $\Bx$. Here, the likelihood $\pLkhd(\By|\Bx)$ is given by the Poisson noise model and $\pprior(\Bx|\BxMR)$ is a prior conditioned on the MR image $\BxMR$, which will be learned via a score model. Using this decomposition, the reverse SDE, given the MR image, can be written as 
\begin{equation}
    \label{eq:cond_reverse_sde_posterior_guided}
    d \Bxt = [\Bf(\Bxt, t) - g(t)^2 \left( \nabla_{\Bx} \log \pLkhdt(\By|\Bx) + \nabla_{\Bx} \log \pprior(\Bx|\BxMR) \right)]  d t + g(t) d \Bar{\mathbf{w}}_t.
\end{equation}
We can use \PETNaive~ or \PETDPS~ to approximate the score of the time dependent likelihood $\nabla_{\Bx} \log \pLkhdt(\By|\Bx)$. However, we have to train a score model, conditioned on the MR image, to estimate the conditional score function
\begin{equation}
    s_\theta(\Bxt; t, \BxMR) \approx \nabla_\Bx \log \ppriort(\Bx | \BxMR),
\end{equation}
where $\BxMR$ is an additional input to the score model. \blue{This method was recently proposed and applied to PET image denoising \citep{gong2022pet}. To train such a score model we need a paired dataset $\{(\Bx^i,\mathbf{x}_{\text{MR}}^i)\}_{i=1}^m$ of PET images and corresponding MR images.} In contrast, using the Classifier Free Guidance (CFG) framework \citep{ho2022classifier}, we only need a partly paired dataset, i.e. besides paired data $\{(\Bx^i,\mathbf{x}_{\text{MR}}^i)\}_{i=1}^{m_1}$ we can also make use of unpaired data $\{\Bx^i\}_{i=1}^{m_2}$. In particular, CFG trains both a conditional and unconditional score model simultaneously and utilises their combination during the sampling process. CFG uses a zero image $\mathbf{0}$ to distinguish between the conditional and unconditional score model
\begin{equation}
    s_\theta(\Bxt; t, \BxMR) \approx \nabla_{\Bx} \log \ppriort(\Bxt | \BxMR) \quad \mbox{and}\quad s_\theta(\Bxt;t, \BxMR = \mathbf{0}) \approx \nabla_{\Bx} \log \ppriort(\Bxt),
\end{equation}
yielding a conditional DSM objective
\begin{equation}
\label{eq:CFG_denoising_score_matching}
    \begin{split}
            \!\E_{t \sim U[0,T]} \E_{\Bxzero, \BxMR \sim \pprior} \E_{\Bxt \sim p_t(\Bxt|\Bxzero)} \E_{\rho \sim B(q)}\!\left[ \omega_t \| s_\theta(\Bxt, t; \rho~ \BxMR) \!-\! \nabla_{\Bx} \log p_t(\Bxt|\Bxzero) \|_2^2 \right]\!\},
    \end{split}
\end{equation}
where $B(q)$ is a Bernoulli distribution with parameter $q$. Thus, if the additional MR input is set to zero, the conditional DSM loss matches the unconditional DSM loss defined in Eq.~\eqref{eq:denoising_score_matching}. After training, CFG defines a combined score model 
\begin{equation}
        \Tilde{s}_\theta(\Bxt;t, \BxMR) = (1+w)s_\theta(\Bxt; t, \BxMR) - w s_\theta(\Bxt; t, \mathbf{0}),
\end{equation}
as a linear combination with $w$ as the guidance strength. This combined score model $s_\theta(\Bxt;t, \BxMR)$ can then be used for any of the presented sampling methods. 

\section{Experimental Setup}
\label{sec:experimental_setup}

\subsection{Dataset and Evaluation Metrics}
\label{sec:dataset}

We use the BrainWeb dataset consisting of $20$ patient-realistic volumes \citep{aubert2006twenty}. The tracer simulated was ${}^{18}$F-Fluorodeoxyglucose (FDG) and the volumes were further perturbed by three realisations of random distortions \citep{georg2021data}. $19$ out of the $20$ volumes were used for training. Axial slices with non-zero intensity were \blue{removed}, resulting in a training dataset of \num{4569} slices. \blue{We conducted two simulations for the evaluation, i.e., image and measurement. The image simulation utilised the segmentation mask of the held-out volume, subject 04. In 2D, an FDG tracer was simulated by perturbing the contrast, applying a bias-field to grey matter, blurring and adding noise as per \cite{georg2021data}. In 3D, FDG and Amyloid tracers were independently simulated, and blurring and noise were added. The Amyloid tracer was included to provide a further OOD evaluation set. To further diversify the evaluation sets, simulated lesions were included in both 3D simulations and added to an additional 2D evaluation set. Lesions were simulated as local regions of ellipsoidal hyper-intensity of random size and location within soft-tissue, and allowed for local evaluation metrics to be computed. The setting of measurement simulation is described below separately for the 2D and 3D cases.}

\blue{For 2D evaluation, resolution modelling, attenuation, sensitivity, and background contamination were modelled and subsequently included in the forward model utilising ParallelProj \citep{schramm2021parallelproj}. We use $20$ equidistant axial slices from the simulation of subject 04}. The noise level of simulated measurements was set by re-scaling forward projected ground truth images, where the scale ensured that the total counts divided by emission volume was 2.5 or 10. These rescaled measurements are the clean measurements, which were then corrupted with Poisson noise and constant background contamination. In addition, $10$ noise realisations were obtained. Herein, we refer to the noise levels as $2.5$ and $10$, where the total true counts averaged over evaluation dataset were \num{122808} and \num{491232}, respectively. 

For the 3D evaluation, measurements of subject 04 were simulated with an Siemens Biograph mMR scanner geometry \citep{Karlberg2016}. Measurements with detector sensitivities and attenuation were simulated and included in the forward model using SIRF and STIR \citep{OvtchinnikovBKP20,thielemans2012stir}. The noise level was equivalent to 40 million counts without background, and 5 noisy realisations were obtained. Unless otherwise specified, the projector and measurements were split into 28 ordered subsets in the experiments below.

We evaluate the performance between reconstructions and ground truths using two global metrics: Peak-Signal-to-Noise Ratio (PSNR) and Structural Similarity Index Measure (SSIM) \citep{wang2004image}.
Moreover, we compute two local quality scores over a Region of Interest (ROI). First, to quantify the detectability of lesions, we compute the Contrast Recovery Coefficient (CRC). 
Second, the noise in reconstructions is estimated over background ROIs using Standard Deviation (STD), which computes standard deviation across realisations and then averages over ROI voxels \citep{Tong2010}. In 2D, we evaluate the reconstruction consistency by computing the Kullback-Leibler Divergence (KLDIV) between measurements $\By$ and estimated measurements $\Bbary = \BA \hat{\Bx} + \Bbarb$, where $\hat{\Bx}$ denotes the reconstruction. Furthermore, we include the ``mean KL'' between the noisy and clean measurements across the 2D evaluation dataset. More information about quality metrics can be found in Appendix \ref{app:eval_metrics}. 

We present tables of best performing methods with optimal penalty strengths, as well as qualitative figures of reconstructed images. Furthermore, to allow direct comparison between methods, we give sensitivity plots of PSNR, SSIM, KLDIV or CRC vs. STDs. Since STD gives an estimate of the noise in the image, these plots can show the effect of varying penalty strength on reconstruction quality or data-consistency. A lower STD typically corresponds to lower data-fidelity (a higher prior strength), and the converse is true for higher STD. In practice, with a generative model as a prior, higher penalty strengths do not necessarily lead to a lower STD as there may be multiple reconstruction with high likelihood under the model. Variations in STD are further exacerbated by approximate nature and stochasticity of SGM sampling.



\subsection{Comparison Methods}
In the 2D setting, we compare against two established supervised learning methods used in medical image reconstruction: the UNet post-processing FBPConvNet \citep{Jin2017fbpconvnet} and unrolled iterative learned primal dual \citep{adler2018learned}. We modify both models for PET reconstruction; the post-processing method is referred to as \OSEMUNet, and the unrolled method as \LearnedPD. Additionally we compare against a state-of-the-art SGM approach for PET image denoising \citep{gong2022pet}, referred to as \NaiveOSEM. This denoising approach replaces the likelihood on the measurements with a likelihood modelled as a Gaussian centred at the noisy reconstruction. Therefore, \NaiveOSEM~is able to use the same pre-trained score model as our proposed \PETNaive{}, \PETDPS{}, and \PETDDS{} methods. 

For 3D evaluation, Deep Image Prior (DIP) reconstruction was included as an unsupervised comparison method with a 3D network architecture well-established in literature \citep{Gong2019,ote2023list,Imraj:fully3D}. For comparison, converged MAP solutions with an RDP regulariser were computed. BSREM algorithm was used with a range of penalty strengths, cf. PET background Sect. \ref{sec:fundamentals_of_pet}. 

Further details on all comparison methods can be found in Appendix \ref{app:baseline_methods}.

\section{Numerical Experiments}
\label{sec:results}

The first set of experiments investigates the performance of the SGM methods (\NaiveOSEM{}, \PETNaive{}, \PETDPS{}, \PETDDS{}) against one-another and against established supervised methods (\OSEMUNet{}, \LearnedPD{}). This is done in 2D and at two noise levels, with and without lesions. In the second set of experiments we present results with MR image guidance. The last set of experiments investigates the best performing SGM method (\PETDDS) on 3D reconstruction, and provides a comparison against classical MAP and state-of-the-art DIP reconstructions with lesions and two simulated tracers. For all SGM results we make use of a single score model trained on the dataset of axial BrainWeb slices described in Section~\ref{sec:dataset}. The details about the training process and network architecture can be found in Appendix~\ref{app:training_SGM}.
Further results can be found in the Appendix~\ref{app:additional_results}. All results were computed with a single NVIDIA GeForce RTX 3090.

\subsection{2D Reconstruction}

The aim of 2D experiments is to benchmark the SGM and supervised methods, and analyse the stability of SGM methods with respect to the choice of different penalty strengths $\lambda_t^\text{\Naive}$, $\lambda_t^\text{DPS}$ and $\lambda^\text{DDS}$. The penalty strengths for \PETNaive{} and \PETDPS{} depends on the time step $t$, and the details about their specific choice can be found in Appendix \ref{app:experimental_details}.


\subsubsection{Reconstruction without Lesion}

\begin{figure}[t]
    \centering
    \vspace{-.8em}
    \includegraphics[width=\textwidth,trim={0 0.25cm 0 0},clip]{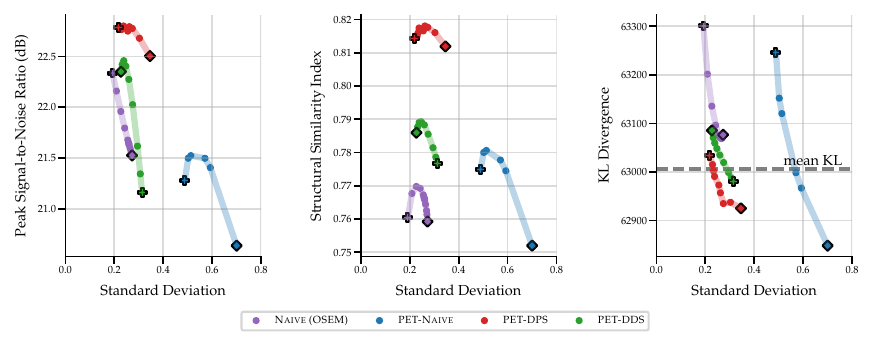}
    \caption{\vspace{-1em}Results for BrainWeb \textbf{without lesions} with noise level \textbf{2.5} for different penalty parameters. Standard deviation is across reconstructions from different realisations of measurements. \blue{The points represent different values of the parameter $\lambda$, and the notation \protect \usym{2719} and $\Diamondblack$ denote the smallest and largest value of $\lambda$, respectively.}}
    \label{fig:brainweb_no_lesion_2.5}
    \vspace{-1em}
\end{figure}

The results in Fig. \ref{fig:brainweb_no_lesion_2.5} show that the performance of the four SGM methods vary greatly for data of noise level 2.5 with no lesions. \PETDPS{} is the best performing method, consistently giving high PSNR, SSIM and low KLDIV values. However, it is also computationally the most expensive, requiring \num{1000} steps with back-propagation through the score model. \PETDDS{} preforms competitively with a much lower computational overhead of \num{100} steps without score model back-propagation. \NaiveOSEM{} performs well with regards to PSNR, but performs poorly in terms of data-consistency (KLDIV) and SSIM. As \NaiveOSEM{} computes the likelihood on an early-stopped OSEM image, increasing data-consistency ensures the reconstruction approaches the OSEM image. The maximum achievable likelihood of \NaiveOSEM{} does not give a KLDIV lower than the ``mean KL''. Hence it is not deemed a strong surrogate to the true likelihood computed on measurements. The \PETNaive{} reconstructions have substantially higher STD values. This is attributed to instability when computing the PLL gradient due to non-negativity projection directly applied on $\Bxt$.

\begin{table}[h!]
\setlength{\tabcolsep}{5pt}
\centering
\caption{\blue{The mean quality score and standard error} using the best hyperparameters for each method for BrainWeb \textbf{without} lesions for noise level \textbf{2.5} \blue{(out-of-distribution)} and \textbf{10} \blue{(in-distribution)}. \blue{The penalty strength used for each SGM method is denoted by $\lambda$.} The best SGM is highlighted in grey, and overall best metric is underlined. \blue{Supervised methods are trained on data with noise levels 5, 10 and 50.} 
\vspace{-1em}}
\begin{tabular}{lll|l|l}
\textbf{ Noise Level} & \textbf{Model} & \textbf{Method} & \textbf{PSNR, $\lambda$} & \textbf{SSIM, $\lambda$}   \\ \midrule
\multirowcell{6}{\textbf{2.5}} &\multirowcell{4}{\textbf{\small Score-based}\\ \textbf{\small Generative}} & \textbf{\small\NaiveOSEM}  & $22.38${\footnotesize $\pm 0.82$}, {\footnotesize $0.527$}  & $0.770${\footnotesize $\pm 0.02$}, {\footnotesize $3.08$}  \\
& & \textbf{\small \PETNaive}  & $21.52${\footnotesize $\pm 0.84$}, {\footnotesize $12.0$}  & $0.781${\footnotesize $\pm 0.01$}, {\footnotesize $12.0$}  \\
& & \textbf{\small \PETDPS} & {\cellcolor[gray]{.9}}$22.80${\footnotesize $\pm 0.81$}, {\footnotesize $650.$}  & {\cellcolor[gray]{.9}}\underline{$0.818$}{\footnotesize $\pm 0.01$}, {\footnotesize $750.$}  \\
& & \textbf{\small \PETDDS} & $22.46${\footnotesize $\pm 0.82$}, {\footnotesize $0.25$}  & $0.789${\footnotesize $\pm 0.02$}, {\footnotesize $0.2$}  \\ \cmidrule{2-5}
&\multirowcell{2}{\textbf{\small Supervised}}& \textbf{\small \LearnedPD} & \underline{$23.06$}{\footnotesize $\pm 0.85$}, {\footnotesize N/A}  & $0.806${\footnotesize $\pm 0.01$}, {\footnotesize N/A}   \\ 
& & \textbf{\small \OSEMUNet} & $22.80${\footnotesize $\pm 0.82$}, {\footnotesize N/A}  & $0.798${\footnotesize $\pm 0.01$}, {\footnotesize N/A}   \\ \midrule 
\multirowcell{6}{\textbf{10}} & \multirowcell{4}{\textbf{\small Score-based}\\ \textbf{\small Generative}}&\textbf{\footnotesize \NaiveOSEM}  & $23.40${\footnotesize $\pm 0.84$}, {\footnotesize $0.2$}  & $0.793${\footnotesize $\pm 0.02$}, {\footnotesize $0.9$}  \\
& &\textbf{\small \PETNaive}  & $22.81${\footnotesize $\pm 0.87$}, {\footnotesize $10.$}  & $0.815${\footnotesize $\pm 0.01$}, {\footnotesize $10.$}  \\
& &\textbf{\small \PETDPS} & {\cellcolor[gray]{.9}}$23.70${\footnotesize $\pm 0.83$}, {\footnotesize $400.$}  & {\cellcolor[gray]{.9}}$0.850${\footnotesize $\pm 0.01$}, {\footnotesize $400.$}  \\
& &\textbf{\small \PETDDS} & $23.55${\footnotesize $\pm 0.77$}, {\footnotesize $0.025$}  & $0.849${\footnotesize $\pm 0.01$}, {\footnotesize $0.025$}  \\ \cmidrule{2-5}
& \multirowcell{2}{\textbf{\small Supervised}} &\textbf{\small \LearnedPD}& \underline{$24.74$}{\footnotesize $\pm 0.91$}, {\footnotesize N/A}  & \underline{$0.861$}{\footnotesize $\pm 0.01$}, {\footnotesize N/A}   \\ 
& &\textbf{\small \small \OSEMUNet}& $24.52${\footnotesize $\pm 0.85$}, {\footnotesize N/A}  & $0.868${\footnotesize $\pm 0.01$}, {\footnotesize N/A}   \\ \bottomrule
\end{tabular}
\label{tab:2d_without_lesion}
\vspace{-1em}
\end{table}

In Table \ref{tab:2d_without_lesion} we show quantitative results of the optimal penalty strength choice for each metric, and comparisons against \OSEMUNet{} and \LearnedPD{}. These supervised methods are trained on data with noise levels of 5, 10 and 50 without lesions. Using noise levels 2.5 and 10 in evaluation allows investigating the effect of OOD noise levels on supervised methods. \LearnedPD{} is the best performing method, giving the best SSIM at noise level 10, and best PSNR at both noise levels. Between noise levels 10 and 2.5 \LearnedPD{} observes a drop of 6.7\%  and 6.6\% for PSNR and SSIM, whereas \PETDPS{} exhibits a drop of 3.4\% and 3.8\%, respectively. 
\PETDPS{} performs competitively across both noise levels and metrics, and gives the best SSIM value at noise level 2.5. \blue{The competitive performance and the reduced performance drop of quality metrics, with increasing noise, provides evidence that \PETDPS{} is more robust to different noise levels.} This may be attributed to the unsupervised nature of SGM methods. Namely, as they are not trained on data of given noise levels they are less affected by distributional differences in noise levels at evaluation and training stages. \blue{However, a more comprehensive study over a wider range of noise levels is needed to robustly support this observation.} \blue{Interestingly, the standard deviation of the different samples is non-negligible in terms of PSNR, but the SSIM is nearly independent of the samples. The larger standard deviation for the PSNR is due to the different dynamic range, e.g., the maximum intensity varies between $11.89$ and $24.15$ for different slices at noise level $2.5$. We leave a full investigation of uncertainty quantification using SGM methods to future work. The supervised methods, i.e., PET-LPD and PET-UNet, perform better than score-based models for both noise levels. Note, this margin is larger for the in-distribution case, i.e., the noise level 10.}

\blue{In the table, we also observe a pronounced change in the selected $\lambda$ values between noise levels 2.5 and 10 for \PETDDS{}, especially when compared with other methods. This phenomenon is attributed to the following fact. In the experiments, the BSREM-like data consistency updates are run for $p$ steps, with a small $p$ (see Appendix~\ref{app:experimental_details} for the heuristic rule to determine $p$), not giving a converged MAP estimate, and there is an implicit regularisation that is proportional to the rate of convergence to noisier images. Hence iterative reconstructions at the noise level 2.5 fit to noise quicker, thereby necessitating much stronger regularisation than the noise level 10.}

\begin{table}[]
\centering
\caption{\blue{The computing time of a single reconstruction, averaged over 5 reconstructions.} \vspace{-1em}}
\begin{tabular}{l|l|l|l}
\textbf{Method} & \textbf{\PETNaive} & \textbf{\PETDPS} & \textbf{\PETDDS} \\ \midrule
\textbf{Time (s)} & 41.52   & 43.64  & 3.90\\\bottomrule
\end{tabular}
\label{tab:petmethods_compared}
\end{table}

\blue{In Table \ref{tab:petmethods_compared}, we compare the computing time for one single reconstruction. \PETNaive{} and \PETDPS{} are largely comparable in terms of inference efficiency, and are about ten times slower than \PETDDS{}. The difference in computing times can be attributed to the fact that \PETDDS{} requires fewer time steps; through the use of the accelerated DDIM sampling method.}

\subsubsection{2D Reconstruction with Lesion}
\begin{figure}[t]
\vspace{-1em}
    \centering
    \includegraphics[width=\textwidth,trim={0 0.25cm 0 0},clip]{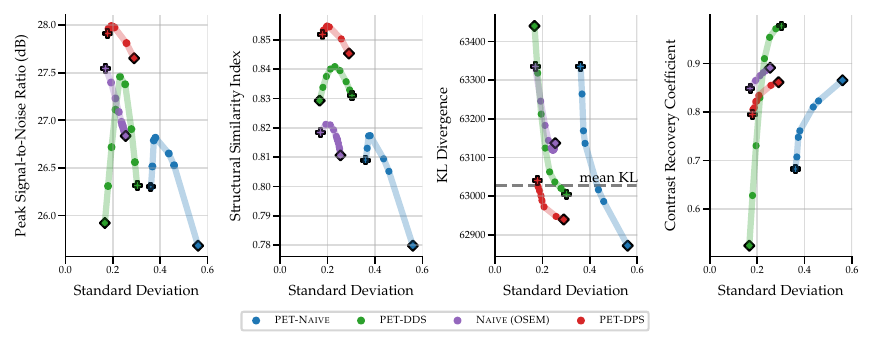}
    \caption{\vspace{-1em}Results for BrainWeb \textbf{with lesions} with noise level \textbf{2.5} for different penalty parameters. Standard deviation is across reconstructions from different realisations of measurements. \blue{The points represent different values of the parameter $\lambda$, and the notation \protect \usym{2719} and $\Diamondblack$ denote the smallest and largest numerical value of $\lambda$, respectively.}}
    \label{fig:brainweb_lesion_2.5}
    \vspace{-1em}
\end{figure}

As the score model was trained on data without lesions, testing on data with simulated hot lesions gives an insight into generalisability to OOD data. The quantitative results in Fig. \ref{fig:brainweb_lesion_2.5} and Table \ref{tab:brainweb_lesion_table} show results that are consistent with those for data with no lesions in Fig. \ref{fig:brainweb_no_lesion_2.5}. CRC was computed to quantify the detectability of hot lesions. The CRC results indicate that \PETDDS{} is better at resolving lesions than other SGM methods. Further, Fig. \ref{fig:brainweb_lesion_2.5} shows a clear trade-off between reconstruction quality in terms of PSNR and SSIM and visibility of lesions. Here, a lower regularisation results in a better performance in terms of CRC. Results for noise level $10$ are shown in Appendix \ref{app:2D_reco}.

Comparing the results between noise levels $2.5$ and $10$ in Table \ref{tab:brainweb_lesion_table}, we observe that SGMs increase CRC values as compared to supervised methods. SGMs also compare favourably with regards to PSNR and SSIM. CRC is local metric that is more relevant than PSNR or SSIM in a clinical setting, as it quantifies the detectability of lesions. Therefore, it is of greater interest to improve this local metric rather than global metrics. With this perspective, SGMs outperform supervised methods, and the best-preforming SGM methods are \PETDPS{} and \PETDDS{} . Due to the performance observed and computational overhead, \PETDDS{} is considered the most appropriate method to test in guided reconstruction and in the 3D setting.

\blue{Numerically, we observe that \PETDDS{} performs better than  \PETDPS{} in terms of CRC but worse in terms of PSNR
and SSIM. Note that PSNR and SSIM are global quality metrics, evaluating the full reconstructed image relative to the reference, whereas CRC is a local quality metric that focuses on the contrast in a specific ROI. While both global and local measures are desirable for PET reconstructions, they are not always consistent. For example, smoothing may suppress the noise and increase PSNR and SSIM, but it can reduce the contrast, leading to a worse CRC. Likewise, if the method preserves details, it may increase the contrast at the expense of having a higher noise in the image and decreasing PSNR and SSIM. Thus, the results in Table \ref{tab:brainweb_lesion_table} indicate that PET-DDS is more effective in enhancing the contrast of the ROIs, but less effective in suppressing the noise, when compared to PET-DPS.}

\begin{table}[h!]
\setlength{\tabcolsep}{5pt}
\centering
\caption{\vspace{-2em}Results using the best hyperparameters for each method for BrainWeb \textbf{with lesions} for noise level \textbf{2.5} and \textbf{10}. \blue{The penalty strength used for each SGM method is denoted by $\lambda$.} The best score-based method is highlighted in grey. The overall best score per noise level is underlined.\vspace{-1em}}
\begin{tabular}{lll|l|l|l}
\textbf{{\small Noise Level}}& \textbf{\small Model} & \textbf{\small Method} & \textbf{\small PSNR, $\lambda$} & \textbf{\small SSIM, $\lambda$} & \textbf{\small CRC, $\lambda$}  \\ \midrule
\multirowcell{6}{\small \textbf{2.5}} &\multirowcell{4}{\textbf{\footnotesize Score-based}\\ \textbf{\footnotesize Generative}} & \textbf{\footnotesize\NaiveOSEM}  & $27.60${\footnotesize $\pm 0.87$}, {\footnotesize $0.527$}  & $0.821${\footnotesize $\pm 0.02$}, {\footnotesize $1.71$} & $0.891${\footnotesize $\pm 0.02$}, {\footnotesize $50.$} \\
& & \textbf{\footnotesize \PETNaive}  & $26.82${\footnotesize $\pm 0.90$}, {\footnotesize$12.$} & $0.817${\footnotesize $\pm 0.02$}, {\footnotesize$12.$} & $0.908${\footnotesize $\pm 0.03$}, {\footnotesize $50.$} \\
& & \textbf{\footnotesize \PETDPS} & {\cellcolor[gray]{.9}}$27.99${\footnotesize $\pm 0.85$}, {\footnotesize$625.$} & {\cellcolor[gray]{.9}}\underline{$0.855$}{\footnotesize $\pm 0.01$}, {\footnotesize$650.$} & $0.886${\footnotesize $\pm 0.02$}, {\footnotesize $1500.$} \\
& & \textbf{\footnotesize \PETDDS} &  $27.46${\footnotesize $\pm 0.83$}, {\footnotesize$0.15$} & $0.841${\footnotesize $\pm 0.01$}, {\footnotesize$0.15$} & {\cellcolor[gray]{.9}}\underline{$0.977$}{\footnotesize $\pm 0.01$}, {\footnotesize $0.01$}\\ \cmidrule{2-6}
&\multirowcell{2}{\textbf{\footnotesize Supervised}}& \textbf{\footnotesize \LearnedPD} &\underline{$28.40$}{\footnotesize $\pm 0.92$}, {\footnotesize N/A} & $0.853${\footnotesize $\pm 0.01$}, {\footnotesize N/A} & $0.865${\footnotesize $\pm 0.03$}, {\footnotesize N/A}  \\ 
& & \textbf{\footnotesize \OSEMUNet} & $27.74${\footnotesize $\pm 0.83$}, {\footnotesize N/A} & $0.836${\footnotesize $\pm 0.01$}, {\footnotesize N/A} & $0.805${\footnotesize $\pm 0.03$}, {\footnotesize N/A}  \\ \midrule 
\multirowcell{6}{\small \textbf{10}} & \multirowcell{4}{\textbf{\footnotesize Score-based}\\ \textbf{\footnotesize Generative}}&\textbf{\footnotesize \NaiveOSEM}  & $28.87${\footnotesize $\pm 0.93$}, {\footnotesize $0.25$}  & $0.847${\footnotesize $\pm 0.01$}, {\footnotesize $0.9$} & $0.902${\footnotesize $\pm 0.02$}, {\footnotesize $4.$} \\
& &\textbf{\footnotesize \PETNaive}  & $28.07${\footnotesize $\pm 0.94$}, {\footnotesize$10.$} & $0.845${\footnotesize $\pm 0.01$}, {\footnotesize$7.5$} & $0.911${\footnotesize $\pm 0.02$}, {\footnotesize $20.$} \\
& &\textbf{\footnotesize \PETDPS} & {\cellcolor[gray]{.9}}$29.01${\footnotesize $\pm 0.87$}, {\footnotesize$400.$} & $0.878${\footnotesize $\pm 0.01$}, {\footnotesize$400.$} & $0.920${\footnotesize $\pm 0.02$}, {\footnotesize $550.$} \\
& &\textbf{\footnotesize \PETDDS} &  $28.99${\footnotesize $\pm 0.88$}, {\footnotesize$0.025$} & {\cellcolor[gray]{.9}}$0.879${\footnotesize $\pm 0.01$}, {\footnotesize$0.025$} & {\cellcolor[gray]{.9}}\underline{$1.00$}{\footnotesize $\pm 0.01$}, {\footnotesize $0.$}\tablefootnote{Regularised due to denoised score estimate initialisation.} \\ \cmidrule{2-6}
& \multirowcell{2}{\textbf{\footnotesize Supervised}} &\textbf{\footnotesize \LearnedPD} &\underline{$30.07$}{\footnotesize $\pm 0.96$}, {\footnotesize N/A} & \underline{$0.894$}{\footnotesize $\pm 0.01$}, {\footnotesize N/A} & $0.904${\footnotesize $\pm 0.02$}, {\footnotesize N/A}  \\ 
& &\textbf{\footnotesize \small \OSEMUNet} &$29.41${\footnotesize $\pm 0.82$}, {\footnotesize N/A} & $0.889${\footnotesize $\pm 0.01$}, {\footnotesize N/A} & $0.865${\footnotesize $\pm 0.03$}, {\footnotesize N/A}  \\ \bottomrule
\end{tabular}
\label{tab:brainweb_lesion_table}
\vspace{-1em}
\end{table}

\subsubsection{MR Guided Reconstruction}

Experiments with and without additional MR image guidance were conducted to illustrate the flexibility of the proposed approach, and tested at three guidance strengths $w = 0.25$, $0.5$, $1.0$, where the guidance strength $w$ closer to zero constitutes more guidance. \blue{We use
high resolution, low noise T1-weighted simulated MR image, which allows testing the worst case scenario where the MR image misrepresents the lesions.} The results with best hyper-parameters are given in Table \ref{tab:brainweb_MR_guided_table_25}. It is observed that there are significant improvements to PSNR ($>18\%$) and SSIM ($>13\%$) with guidance. On PET data with lesions, the lesions were only simulated for PET and not MR images. Therefore, the data was of a worse-case scenario where clinically important features are only present in the PET image. The results with lesions show increasing the guidance strength decreased of CRC values and the lesions were more difficult to detect - see Fig. \ref{fig:2D_guided_img_ext_2}. Conversely, the PSNR and SSIM values on with lesions data increased with $w$ closer to zero (more guidance). This highlights the potential dangers of guidance, as well as the importance of evaluating local and global quality metrics.

\begin{table}[h!]
\centering
\caption{\vspace{-1.5em}Results using the best hyperparameters for SGM methods for noise level \textbf{2.5} with \textbf{MR image guidance}. \blue{The penalty strength used for each SGM method is denoted by $\lambda$.} The best method by performance metric is highlighted in grey for with/without lesion. The penalty strength is tuned for each method individually.\vspace{-1em}}
\begin{tabular}{l|l|l|l|l|l}
& \multicolumn{2}{c|}{\textbf{without lesions}} & \multicolumn{3}{c}{\textbf{with lesions}} \\ 
  & \textbf{PSNR, $\lambda$} & \textbf{SSIM, $\lambda$} & \textbf{PSNR, $\lambda$} & \textbf{SSIM, $\lambda$} & \textbf{CRC, $\lambda$}  \\ \midrule
\textbf{DDS} (w/o MR) & $22.46$, {\footnotesize $0.25$}  & $0.789$, {\footnotesize $0.2$} & $27.46$, {\footnotesize $0.15$}  & $0.841$, {\footnotesize $0.15$} & $0.910$, {\footnotesize $0.01$} \\
\textbf{DDS $w=0.25$}  & {\cellcolor[gray]{.9}}$30.22$, {\footnotesize $0.35$}  & {\cellcolor[gray]{.9}}$0.950$, {\footnotesize $0.35$} & {\cellcolor[gray]{.9}}$31.21$, {\footnotesize $0.15$}  & {\cellcolor[gray]{.9}}$0.954$, {\footnotesize $0.25$} & $0.726$, {\footnotesize $0.0$}\\
\textbf{DDS $w=0.5$} & $29.32$, {\footnotesize $0.25$}  & $0.940$, {\footnotesize $0.25$}& $31.12$, {\footnotesize $0.15$}  & $0.946$, {\footnotesize $0.25$} & $0.778$, {\footnotesize $0.0$} \\
\textbf{DDS $w=1.0$} & $26.66$, {\footnotesize $0.15$}  & $0.899$, {\footnotesize $0.15$} & $29.31$, {\footnotesize $0.1$}  & $0.906$, {\footnotesize $0.15$} & {\cellcolor[gray]{.9}}$0.939$, {\footnotesize $0.0$}\\ \bottomrule
\end{tabular}
\label{tab:brainweb_MR_guided_table_25}
\vspace{-.5em}
\end{table}

\begin{figure}[t]
    \centering
  \includegraphics[width=1.\linewidth]{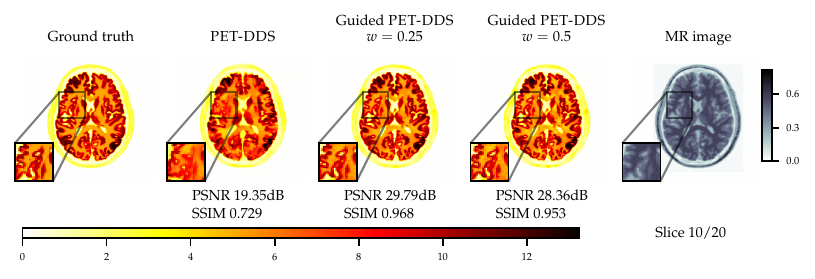}\\

\includegraphics[clip, trim=0.cm 0.cm 0.cm 2.1em, width=1.\linewidth]{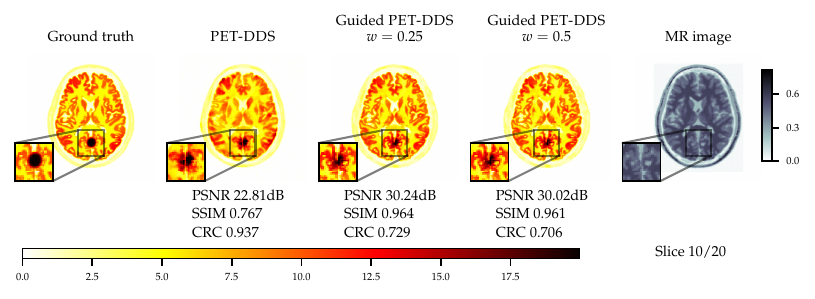}
    \caption{\vspace{-1em}Comparisons of single slice reconstructions with the \PETDDS~MR guided vs. unguided at noise level \blue{\textbf{2.5 without lesion} (top) and \textbf{with lesion} (bottom).}}
    \label{fig:2D_guided_img}
    \vspace{-1.5em}
\end{figure}

From Fig. \ref{fig:2D_guided_img} a reconstruction without guidance and with guidance of various strengths is presented for data without and with lesions at noise level 2.5. The reconstructions indicate that MR guidance helps to reconstruct the specific anatomical boundaries and structure, i.e., white matter tracts. In Appendix \ref{app:mr_guidance_results} we give additional qualitative slices with and without lesions, cf. Figs. \ref{fig:2D_guided_img_ext_1} and \ref{fig:2D_guided_img_ext_2}, and the associated sensitivity plots in Figs. \ref{fig:results_no_tumour_25_guided_sweep} and \ref{fig:results_tumour_25_guided_sweep}.

\subsection{3D Reconstruction}

\begin{figure}[t]
    \centering
    \includegraphics[width=1.\textwidth,trim={0 0.25cm 0 0},clip]{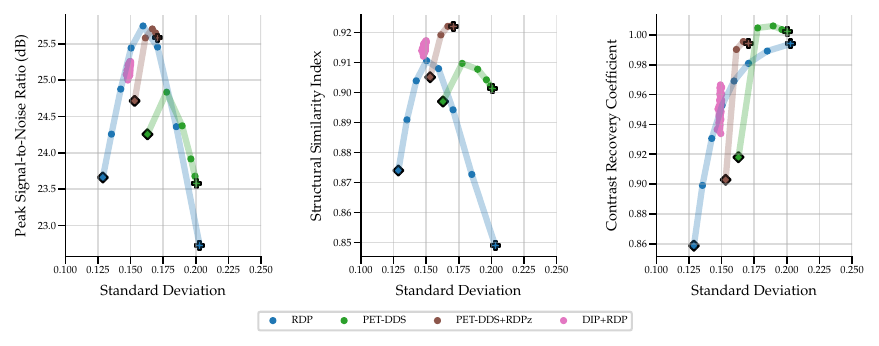}
    \caption{\vspace{-1em}Results for 3D reconstruction using the FDG tracer for different penalty values. PET-DDS-RDP$_z$ $\beta = 21.9$, and DIP+RDP $\beta=0.1$. Standard deviation is across reconstructions from different realisations of measurements. \blue{For DIP, the points corresponds to various number of optimisation steps. For the other methods, the points represent different values of the parameter $\lambda$, and the notation \protect \usym{2719} and $\Diamondblack$ denote the smallest and largest numerical value of $\lambda$, respectively.}
    }
    \label{fig:results_3d_fdg}
\end{figure}

\begin{figure}[h!]
    \centering
    \includegraphics[width=1\textwidth]{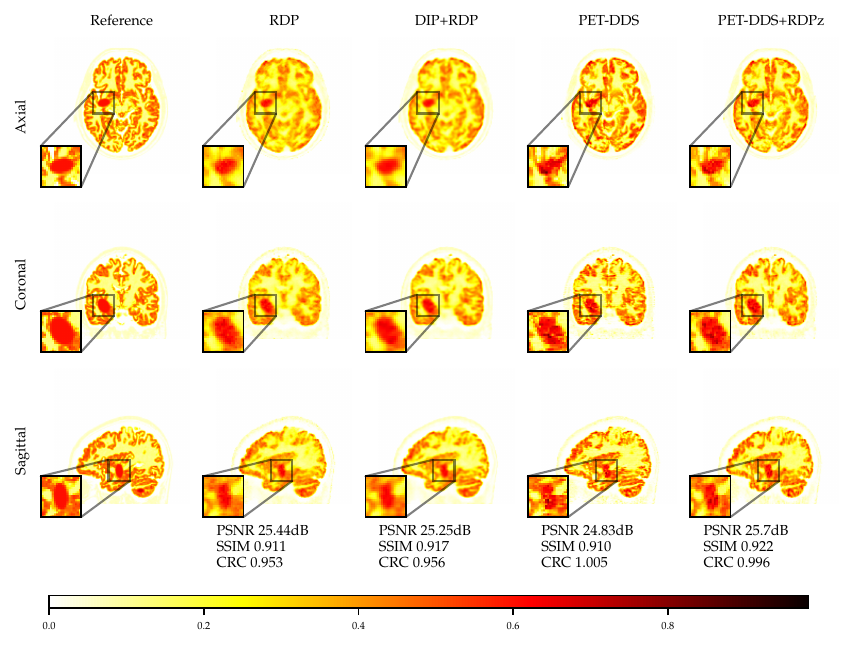}

    \caption{\vspace{-1em}3D reconstruction for the different methods with FDG tracer, and metrics computed on the inset lesion.}
    \label{fig:results_3d_fdq_images}
\vspace{-1em}
\end{figure}

Full 3D reconstructions were analysed for two tracers with simulated lesions. We evaluate the performance of \PETDDS{} with additional RDP regularisation in the $z$-direction perpendicular to axial slices (termed RDPz), and introduce subset-based data consistency updates as in Eq.~\eqref{eq:pet_dds_map}. Acceleration of \PETDDS{} was obtained through the use of subset-based data consistency updates, see Table \ref{tab:subsettime}. For further experiments 28 subsets were used. We compare against a BSREM computed MAP solution with RDP, and DIP with RDP, similar to \cite{Imraj:fully3D}. In Fig. \ref{fig:results_3d_fdg} we show sensitivity plots for the FDG tracer and in Fig. \ref{fig:results_3d_fdq_images} we plot the axial, coronal and sagittal slices centred on the lesion location. Additionally, sensitivity curves for the amyloid tracer are given in Fig. \ref{fig:results_3d_amyloid}, and the associated reconstructions are available in Appendix \ref{app:additional_results}, see Fig. \ref{fig:results_3d_amyloid_images}. \blue{In Table \ref{tab:subsettime}, we present the computing time for 3D \PETDDS{}+RDPz versus the number of subsets. The results indicate that the computing time decreases with the number of subsets while having minimal affect on evaluation metrics, indicating the preference for using more subsets for better computational efficiency in practice.}

\begin{table}[h!]
\vspace{-1em}
\centering
\caption{\blue{3D \PETDDS{}+RDPz computing time with different numbers of subsets. Quality metrics computed on the first realisation of FDG tracer measurements using 3D \PETDDS{}+RDPz with $\lambda=158.0$ and $\beta=21.9$. Best values are highlighted in grey.}}
\begin{tabular}{l|l|l|l|l|l|l}
\textbf{Number of subsets}  & \textbf{1}&\textbf{4}&\textbf{7}&\textbf{14}&\textbf{28}&\textbf{42}  \\\midrule
\textbf{Reconstruction time (min)}  & $47.8$ & $13.6$ & $8.6$ & $5.1$ & $3.4$ & {\cellcolor[gray]{.9}}$2.8$ \\ 
\textbf{PSNR} & {\cellcolor[gray]{.9}}$25.91$& $25.90$ & $25.89$ & $25.84$& $25.72$&$25.57$ \\ 
\textbf{SSIM} & {\cellcolor[gray]{.9}}$0.927$ & $0.927$ & $0.927$ & $0.925$& $0.922$ & $0.919$\\ 
\textbf{CRC} & {\cellcolor[gray]{.9}}$0.990$ & $0.990$ & $0.990$ & $0.985$ & $0.987$ & $0.988$ \\
\bottomrule
\end{tabular}
\label{tab:subsettime}
\vspace{-.5em}
\end{table}

The FDG tracer sensitivity plot in Fig. \ref{tab:brainweb_3D} shows that adding RDPz into \PETDDS{} improves SSIM and CRC metrics, while classical RDP provides highest PSNR values. 
Since PSNR is computed using a mean squared reconstruction error, the resulting metric is biased toward blurrier reconstructions. This can be observed in the qualitative images given in Fig. \ref{fig:results_3d_fdq_images}, where RDP gives high PSNR values while the image insets show excessive blurring on the lesion. \PETDDS{} without RDPz performs worse than with RDPz, since the score model only acts on axial slices and, without RDPz, consistency in $z$-direction is only ensured through data consistency. Qualitatively this can be observed in Fig. \ref{fig:results_3d_fdq_images}, where coronal and sagittal slices display discontinuities in the $z$-direction whereas the axial slice is smoother. DIP reconstructions give improvements in SSIM and CRC as compared to classical RDP results, but fail to improve PSNR. Results with OOD Amyloid tracer show milder improvements with \PETDDS{}, with trends similar to those seen with the FDG tracer.

\blue{PET reconstructions by the proposed methods in Fig. \ref{fig:results_3d_fdq_images} closely match the ground truth. However, unlike in RDP and DIP-RDP reconstructions, the results by \PETDDS{} and \PETDDS-RDPz{} exhibit regions with much higher intensity values than in neighbouring areas. The inhomogeneity is attributed to the presence of noise in the data, and PET-DDS without RDPz is expected to be more inhomogenous for slices that are not axial, because the score-model was only enforced axially.}

\begin{table}[b]
\centering
\caption{\vspace{-1em}Results using the best hyperparameters for each method for 3D BrainWeb data with FDG and Amyloid tracers.\blue{ The penalty strength used for each SGM method is denoted by $\lambda$. The best performing method is highlighted in grey.}
\vspace{-1em}}
\begin{tabular}{ll|lll}
\textbf{Tracer} & \textbf{Method} & \textbf{PSNR, $\lambda$} & \textbf{SSIM, $\lambda$} & \textbf{CRC, $\lambda$}  \\ \midrule
\multirowcell{4}{\textbf{FDG}}&
\textbf{RDP}  & {\cellcolor[gray]{.9}}$25.74$, {\footnotesize $1.81$}  & $0.911$, {\footnotesize $2.77$} & $0.994$, {\footnotesize $0.5$} \\ &
\textbf{DIP+RDP}  & $25.26$, {\footnotesize$9,800$} & $0.917$, {\footnotesize$10,800$} & $0.966$, {\footnotesize $9,500$} \\ &
\textbf{\PETDDS} & $24.83$, {\footnotesize$398$} & $0.910$, {\footnotesize$398$} & {\cellcolor[gray]{.9}}$1.01$, {\footnotesize $158$} \\ &
\textbf{\PETDDS+RDPz} &  $25.70$, {\footnotesize$158$} & {\cellcolor[gray]{.9}}$0.922$, {\footnotesize$63.1$} & $0.996$, {\footnotesize $158$} \\ \midrule
\multirowcell{4}{\textbf{Amyloid}}&
\textbf{RDP}  & {\cellcolor[gray]{.9}}$24.15$, {\footnotesize$2.77$}  & $0.898$, {\footnotesize $1.81$} & $0.996$, {\footnotesize $0.5$} \\ &
\textbf{DIP+RDP}  & $24.10$, {\footnotesize$10,200$} & $0.894$, {\footnotesize$10,800$} & $0.964$, {\footnotesize $9,500$} \\ &
\textbf{\PETDDS} & $23.08$, {\footnotesize$1000$} & $0.890$, {\footnotesize$398$} & {\cellcolor[gray]{.9}}$1.009$, {\footnotesize $10$} \\ &
\textbf{\PETDDS+RDPz} &  {\cellcolor[gray]{.9}}$24.15$, {\footnotesize $398$} & {\cellcolor[gray]{.9}}$0.906$, {\footnotesize$158$} & $0.999$, {\footnotesize $10$} \\
\bottomrule
\end{tabular}
\label{tab:brainweb_3D}
\end{table}

\begin{figure}[h!]
\vspace{-1em}
    \centering
    \includegraphics[width=1.\textwidth,trim={0 0.25cm 0 0},clip]{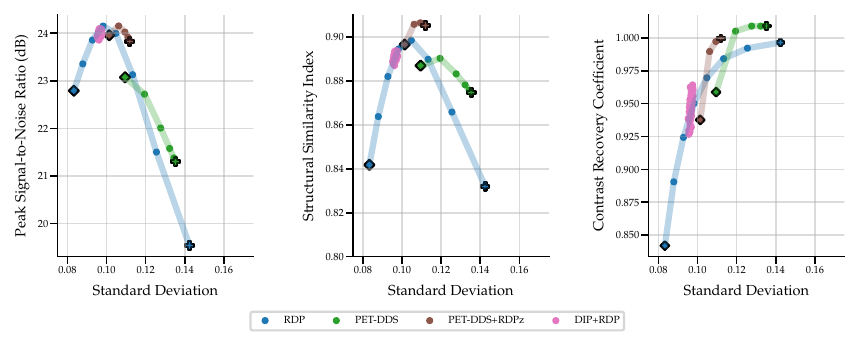}
    \caption{\vspace{-1em}Results for 3D reconstruction using the Amyloid tracer for different penalty values. PET-DDS-RDP$_z$ $\beta = 21.9$, and DIP+RDP $\beta=0.1$. Standard deviation is across reconstructions from different realisations of measurements. \blue{For DIP, the points corresponds to various number of optimisation steps. For the other methods, the points represent different values of the parameter $\lambda$, and the notation \protect \usym{2719} and $\Diamondblack$ denote the smallest and largest numerical value of $\lambda$, respectively.}}
    \label{fig:results_3d_amyloid}
\vspace{-1em}
\end{figure}

\section{Conclusion}
In this work we adapt SGMs for PET image reconstruction by incorporating PET specific constraints, e.g. Poisson noise and non-negativity, into several popular sampling techniques. We further introduce a measurement-based normalisation technique, to improve the generalisability to different intensity values by stabilising the dynamic range encountered by the score model. In future work, reflected SGMs, recently proposed by \cite{lou2023reflected}, could be leveraged to introduce non-negativity into the sampling procedure in a more principled manner.
This work provides a first investigation of the generalisation capabilities by training the score model on patient-realistic slices without lesions and testing on slices with lesions. However, further work is needed to comprehensively evaluate the generalisation performance on \blue{more diverse datasets of} \textit{in-vivo} data, \blue{to} investigate the biases of SGMs, which is vitally important for clinical adoption. 
The proposed SGM sampling methods can produce multiple samples from the posterior $p(\Bx|\By)$, and in this vein one can draw multiple samples from the posterior for empirical uncertainty estimation; this is left for future work.
This work proposes guided SGM reconstruction with an additional MR guidance image using CFG. The preliminary results are promising and further validation is required. A clinically pertinent investigation into robustness to misregistration of the MR image could be investigated. Furthermore, guidance could be extended to a joint PET-MRI reconstruction. Recently, \cite{levac2023mri} used similar ideas for a joint reconstruction of multi-contrast MR images.


\acks{I.R.D. Singh and R. Barbano are supported by the EPSRC-funded UCL Centre for Doctoral Training in Intelligent, Integrated Imaging in Healthcare (i4Health) (EP/S021930/1) and the Department of Health’s NIHR-funded Biomedical Research Centre at University College London Hospitals. A. Denker acknowledges the support by the Deutsche Forschungsgemeinschaft (DFG, German Research Foundation) - Project number 281474342/GRK2224/2. \v{Z}. Kereta was supported by the UK EPSRC grant EP/X010740/1. B. Jin and S. Arridge are supported by the UK EPSRC EP/V026259/1, and B. Jin is also supported by a start-up fund from The Chinese University of Hong Kong. Software used in this project is partially maintained by CCP SyneRBI (EPSRC EP/T026693/1).
P. Maass acknowledges support by DFG-NSFC project M-0187 of the Sino-German Center mobility programme. The authors thank Georg Schramm for help with ParallelProj.
}

%
\ethics{The work follows appropriate ethical standards in conducting research and writing the manuscript, following all applicable laws and regulations regarding treatment of animals or human subjects.}

\coi{We declare we don't have conflicts of interest.}

\bibliography{references}


\clearpage
\appendix
\section{Appendix}

\subsection{Training of Score-based Models}
\label{app:training_SGM}

\paragraph{Forward SDE}

In our experiments, we make use of the variance preserving SDE \citep{ho2020denoising}
\begin{align}
    d \Bxt = - \frac{\beta(t)}{2} \Bxt d t + \sqrt{\beta(t)} d \mathbf{w},
\end{align}
were we employ $\beta(t) = \betamin + t(\betamax - \betamin)$ as a linear schedule with $\betamin = 0.1$ and $\betamax=10$ with a terminal time $T=1$. The coefficients were chosen such that the terminal distribution  approximates a Gaussian, i.e. $p_1(\Bx) \approx \mathcal{N}(0,I)$. We also tested the Variance Exploding (VE) SDE \citep{song2020score}; it was found that \textbf{VE-SDE was more unstable than VP-SDE} for PET image reconstruction. The transition kernel for the variance preserving SDE is a Gaussian, i.e. $p_t(\Bxt|\Bxzero) = \mathcal{N}(\Bxt; \gamma_t \Bxzero, \nu_t^2 I)$, with coefficients
\begin{align}\label{eqn:gammanu_t}
    \gamma_t = \exp{\left(-\frac{1}{2} \int_0^t \beta(s) d s\right)}, \quad \nu_t^2 = 1 - \exp{\left(-\int_0^t \beta(s) d s\right)}.
\end{align}
Using this closed form expression for the transition kernel, the denoising score matching loss can be rewritten as 
\begin{align}
    L_\text{DSM}(\theta) = \mathbb{E}_{t \sim U[0,1]} \mathbb{E}_{\Bxzero \sim \pprior} \mathbb{E}_{\Bz \sim \mathcal{N}(0,I)} \left[\omega_t \left\| s_\theta(\Bxt, t) + \frac{1}{\nu_t}\Bz \right\|_2^2 \right],
\end{align}
with $\Bx_t = \gamma_t \Bx_0 + \nu_t \Bz$. The weighting $\omega_t$ is chosen as $\omega_t = \nu_t^2$ to approximate maximum likelihood training \citep{song2021maximum}.

\paragraph{Model Architecture}
We use the architecture proposed by \cite{dhariwal2021diffusion}\footnote{available at \url{https://github.com/openai/guided-diffusion}}. The architecture is based on the popular U-Net architecture \citep{ronneberger2015u} consisting of a decoder implemented as a stack of residual blocks and downsampling operations and an encoder of residual blocks and upsampling operations. At the lowest resolution ($8 \times 8$), additional global attention layers are used. To incorporate the timestep into each residual block, the authors use adaptive group normalisation (AdaGN) layers defined as $\text{AdaGN}(h, e) = e_s \text{GroupNorm}(h) + e_b$, where $h$ are intermediate features and $e=[e_s, e_b]$ is the encoded time step. The specific implementation and the choice of our hyperparameters can be seen in \href{https://github.com/Imraj-Singh/Score-Based-Generative-Models-for-PET-Image-Reconstruction}{our github}. For the MRI guided model we apply the clean MRI image as an additional channel to the input of the network.

\subsection{Experimental Details}
\label{app:experimental_details}
The sampling methods presented in Section \ref{sec:appl_to_PET} use different penalty strengths in order to scale the likelihood term for \PETNaive~and \PETDPS~or to set the strength of the additional Tikhonov regularisation for \PETDDS. For \Naive~it is recommended to choose $\lambda_t^\text{naive}$ s.t. the penalty is zero at the start of sampling and increased as $t \to 0$ \citep{jalal2021robust}.  We use $\lambda_t = \lambda (1-t)$ in all our experiments. For the \PETDPS~approach \citep{chung2023noise} define the sampling iteration as 
\begin{equation}
\label{eq:dps_sampling}
\begin{split}    
    &\Tilde{\Bx}_{t_{k-1}} = \Bx_{t_k} + [\Bf(\Bx_{t_k}, t_k) - g(t_k)^2 s_\theta(\Bx_{t_k}, t_k)] \Delta t + g(t_k) \sqrt{|\Delta t|} \Bz \quad \Bz \sim \mathcal{N}(0, I),  \\ 
    &\Bx_{t_{k-1}} = \Tilde{\Bx}_{t_{k-1}} - \lambda_{t_k}^\text{DPS} \nabla_\Bx L(\By | \Tweedie(\Bx_{t_k})).
\end{split}
\end{equation}
This is equivalent to the classical Euler-Maruyama scheme, when $\lambda_{t_k}^\text{DPS}$ is chosen in such a way that it incorporates the step size $\Delta t$ and the diffusion function $g(t_k)^2$. We follow \cite{chung2023noise} and define $\lambda_t^\text{DPS} = \frac{\lambda}{D_{KL}(A \Tweedie || \By)}$. For \PETDDS~a constant penalty $\lambda^\text{DDS}$, without time dependency, is used. Heuristically, the number of iterations used for data-consistency projection were adjusted such that the results, with $\lambda^\text{DDS} = 0$, overfit to noise. The penalty strength $\lambda^\text{DDS}$ was then increased to regularise the reconstruction more. In 2D the number of projection steps for \PETDDS{} were set to $4$ for noise level $2.5$ and $15$ for noise level $10$. \blue{In 3D the number of projection steps for \PETDDS{} were set to $5$ for both tracers. For guided reconstruction, $10$ steps for noise level $2.5$ and $20$ steps for noise level $10$ were used for \PETDDS{}.}

\subsection{Baseline Methods}
\label{app:baseline_methods}
\paragraph{Classical Methods}
Relative Difference Prior (RDP) is a common penalty for PET reconstruction \citep{nuyts2002concave}, defined by
\begin{align*}
    R(\Bx) = -\sum_{j=1}^n \sum_{k\in N_j} \frac{(x_j-x_k)^2}{x_j+x_k+\xi\lvert x_j-x_k\rvert},
\end{align*}
where $N_j$ is a pre-defined neighbourhood around $x_j$, typically $3\times3$ in 2D or $3\times3\times3$ in 3D.
$R_z(\Bx)$ is a variant of RDP whereby the  neighbourhood is defined in the axial dimensions in 3D, i.e. $3\times1\times1$. A Neumann boundary was chosen where neighbourhoods that were outside of the domain. The tunable parameter $\xi>0$ controls the degree of edge-preservation ($\xi=1$, in-line with clinical practice), and with its gradient given by 
\begin{align}
    \frac{\partial R(\Bx)}{\partial x_j} = \sum_{k\in N_j} - \frac{(r_{jk}-1)(\xi |r_{jk}-1| + r_{jk} +3)}{(r_{jk}+1 + \xi | r_{jk} - 1|)^2}, \quad \text{with } r_{jk} := \frac{x_{j}}{x_{k}}.
\end{align}
The penalisation is scale-invariant since the gradient is computed using the ratio of voxel values $r_{jk}$. This partially overcomes the issue with the wide dynamic range observed in emission tomography images. For BSREM algorithm the convergence criteria was set based on the change of voxel values within the reconstruction between iterates. Specifically, the change in mean voxel values across non-zero voxel values was less than $0.01\%$, we set the relaxation coefficient to $\zeta = 0.1$.

\paragraph{PET Image Denoising with SGM}
In PET image denoising, the goal is to sample from the posterior $p(\Bx| \mathbf{x}_\text{noisy})$ of the true image $\Bx$ given an initial (low-count) reconstruction $\mathbf{x}_\text{noisy}$. This is differs from PET reconstruction, where the goal is to sample from the posterior $\pPost(\Bx | \By)$ conditioned on the measurements $\By$. In this framework the denoising likelihood is given by Gaussian noise, i.e. 
\begin{align}
    p(\BxOSEM | \Bx) = \mathcal{N}(\mathbf{x}_\text{noisy};\Bx, \sigma_d^2 I),
\end{align}
with the noise level $\sigma_d$ to be specified. Using the \Naive~approximation, we get the following reverse SDE for the PET denoising likelihood
\begin{align}
    \label{eq:naive_pet_denoising}
    d \Bxt = [\Bf(\Bxt, t) - g(t)^2 ( s_\theta(\Bxt, t) - 1/\sigma_d^2 (\mathbf{x}_\text{noisy} - \Bxt)]  d t + g(t) d \Bar{\mathbf{w}}_t.
\end{align}
In our implementation we estimate the initial reconstruction using OSEM with 34 subsets and iterations (i.e. 1 epoch). The same score model $s_\theta(\Bxt, t)$ is used for both PET denoising and reconstruction. The noise level $\sigma_d$ is chosen based on a held-out evaluation dataset.

\paragraph{Supervised Learning}
We are using two popular supervised learning techniques: post-processing and learned iterative methods. For the post-processing method we used a variant of the FBPConvNet \citep{Jin2017fbpconvnet}, modified to PET reconstruction. The input to the FBPConvNet was changed to an OSEM with 34 subsets and iterations, this variation is denoted by \OSEMUNet. For the learned iterative method, we adopt Learned Primal Dual (LPD) \citep{adler2018learned}, referred to as \LearnedPD. For \LearnedPD\, we use the same OSEM reconstruction as initialisation for the primal channels and include the affine forward model with sample specific attenuation maps. Note that these sample specific factors were not included in previous implementation of learned iterative methods for PET image reconstruction \citep{guazzo2021learned}. Three primal and dual unrolled iterations were used. Both of these networks were implemented using Div$\alpha$L \citep{leuschner_johannes_2021_4428220} with only minimal changes to the architecture; \OSEMUNet\, was a UNet with \num{1783249} parameters, and \LearnedPD\, used a block of convolutional filters for each primal and dual network with a total of \num{132300} parameters. Both networks were trained using the dataset in Section \ref{sec:dataset} without lesions and noise levels of $5$, $10$, and $50$. The dataset was split into training and evaluation, and training was terminated when over-fitting was observed. Additionally, data-corrected mean normalisation was included to promote generalisability between noise levels. The code for these supervised learning models is publicly available at \url{https://github.com/Imraj-Singh/pet_supervised_normalisation} \blue{and for further details see \citep{singh2023investigating}}.
\paragraph{Deep Image Prior}
The Deep Image Prior (DIP) \citep{ulyanov2018deep} is a popular framework for unsupervised image reconstruction, relying only on a single measurement. A common problem of the DIP is its tendency to overfit to noise. Therefore some regularisation has to be used. We included RDP into the objective function to elevate the need for early-stopping and prevent over-fitting to noise. The architecture used was a three-scale U-Net \citep{ronneberger2015u} with \num{1606899} parameters, with a rectified linear unit on the output to ensure non-negativity. This architecture is minimally changed from previous applications of DIP to PET \citep{Gong2019,ote2023list,Imraj:fully3D}. DIP results are computed on reconstructions along the optimisation trajectory, every 100 iterations from 6,600 iterations to 11,600.

\subsection{Evaluation metrics}
\label{app:eval_metrics}
In addition to peak-signal-to-noise ratio (PSNR) and structural similarity index measure (SSIM) \citep{wang2004image}, we compute two local quality scores over a Region of Interest (ROI). For reconstructions with lesions a Contrast Recovery Coefficient (CRC) was computed to quantify detectability of these local features. This was computed between lesion $L$ and background $B$ ROIs, these have $N_L$ and $N_B$ number of elements respectively\footnote{We use $L$ to denote the lesion ROI in this section only; in the main manuscript $L$ is the likelihood.}. Additionally, there are $R$ realisations of the measured data. Given an ROI $Z$, we include subscript indices for element and realisation $Z_{r,k}$, where $r$ is the realisation index, and $k$ is the element index. An average over the elements of the ROI is denoted as $\Bar{Z}_r = \frac{1}{N_Z}\sum_{k=1}^{N_Z} Z_{r,k}$. The CRC is defined by
\begin{equation}\label{eqn:metrics_crc}
    \mathrm{CRC} :=     \sum_{r=1}^{R} \left(\frac{\Bar{L}_r}{\Bar{B}_r} - 1 \right)/\left(\frac{L_{\mathrm{t}}}{B_{\mathrm{t}}} - 1\right),
\end{equation}
\vspace{-0.2em}where the subscript $\rm t$ denotes the ground truth ROIs. We study the noise over realisations of the measured data using normalised STD (also referred to as ensemble noise, see \cite{Tong2010}, and is reported to give a true estimate of noise in the image). We define an average over realisations of the ROI as $\Bar{Z}_k = \frac{1}{R}\sum_{r=1}^{R} Z_{r,k}$ where STD is computed on background ROIs it is given by
\begin{equation}\label{eqn:metrics_std}
    \mathrm{STD} := \frac{1}{N_B}\sum_{k=1}^{N_B} \sqrt{\frac{1}{R-1} \sum^{R}_{r=1} \frac{(B_{r,k} - \Bar{B}_k)^2}{\Bar{B}_k}}.
\end{equation}
For reconstructions with lesions the background ROI was used, and without lesions a background of the whole emission volume (defined on reference images) was used. In 2D $R=10$ noise realisations of acquisition data were used, and $R=5$ in 3D. To evaluate the consistency of our reconstructions to the true measurements, we compute  the Kullback-Leibler divergence (KLDIV) 
\begin{align}\label{eqn:metrics_kldiv}
    \mathrm{KLDIV} := \sum_{j=1}^m \bar{y}_j \log\left(\frac{\bar{y}_j}{y_j}\right) - \bar{y}_j + y_j,  
\end{align}
between measurements $\By$ and estimated measurements $\Bbary = \BA \hat{\Bx} + \Bbarb$ where $\hat{\Bx}$ denotes the reconstruction.

\section{Additional Results}\label{app:additional_results}

\subsection{2D Reconstruction}
\label{app:2D_reco}

\begin{figure}[h]
    \centering
    \includegraphics[width=\textwidth]{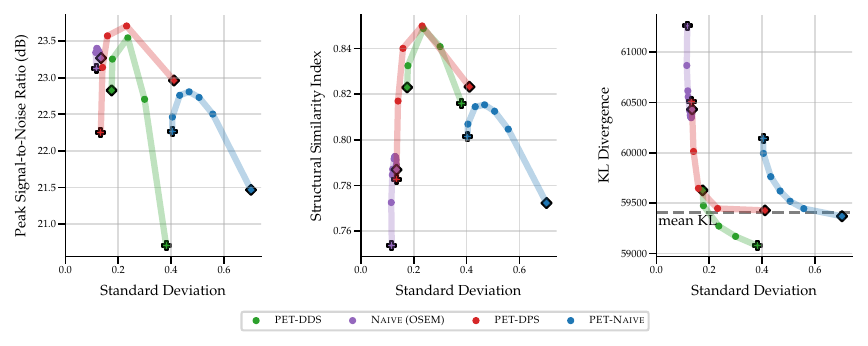}
    \caption{Results for BrainWeb \textbf{without lesions} with noise level \textbf{10} for different penalty parameters. The Standard Deviation is computed over reconstructions of different noise realisations $\By$. \blue{The points represent different values of the parameter $\lambda$, and the notation \protect \usym{2719} and $\Diamondblack$ denote the smallest and largest numerical value of $\lambda$, respectively.} }
    \label{fig:brainweb_no_lesion_10}
\end{figure}

\begin{figure}
    \centering
    \includegraphics[width=\textwidth]{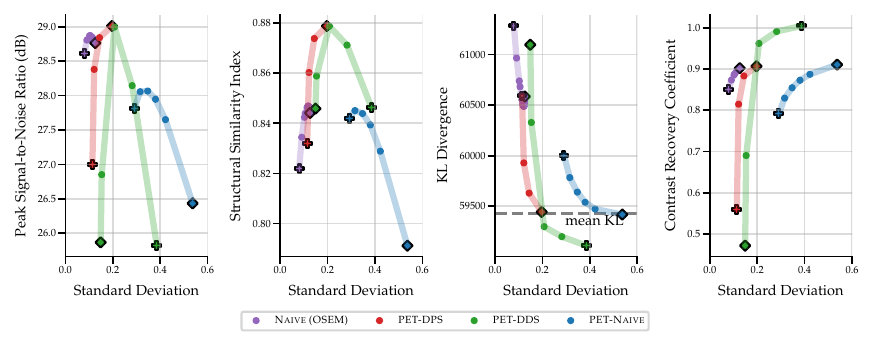}
    \caption{Results for BrainWeb \textbf{with lesions} with noise level \textbf{10} for different penalty parameters. The Standard Deviation is computed over reconstructions of different noise realisations $\By$. \blue{The points represent different values of the parameter $\lambda$, and the notation \protect \usym{2719} and $\Diamondblack$ denote the smallest and largest numerical value of $\lambda$, respectively.}}
    \label{fig:brainweb_lesion_10}
\end{figure}

We show additional sensitivity plots for 2D reconstruction. For noise level $10$ these results are presented in Fig. \ref{fig:brainweb_no_lesion_10} and Fig. \ref{fig:brainweb_lesion_10} without and with lesions, respectively. The results are similar to the settings for noise level $2.5$, as we see a clear trade-off between reconstruction quality in terms of PSNR/SSIM and visibility of lesions in terms of CRC in Fig. \ref{fig:brainweb_lesion_10}. Here, a higher regularisation leads to better PSNR/SSIM scores and a lower regularisation to a better recovery of lesions. A high regularisation, i.e. a high influence of the score model, may lead to a worse reconstruction of the lesions, as the score model was trained on images without lesions.

\subsection{MR guidance} 
\label{app:mr_guidance_results}

\begin{figure}[h!]
    \centering
    \includegraphics[width=1.\textwidth]{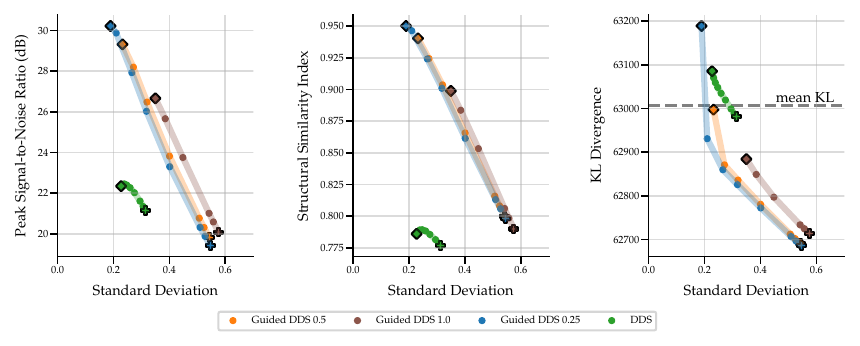}
    \caption{Results for 2D reconstruction guided vs unguided \textbf{without lesions} for noise level \textbf{2.5}. \blue{The points represent different values of the parameter $\lambda$, and the notation \protect \usym{2719} and $\Diamondblack$ denote the smallest and largest numerical value of $\lambda$, respectively.}}
    \label{fig:results_no_tumour_25_guided_sweep}
\end{figure}

\begin{figure}[h!]
    \centering
    \includegraphics[width=1.\textwidth]{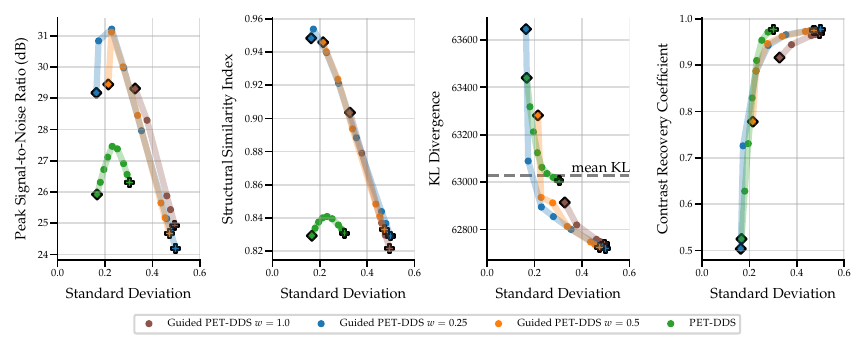}
    \caption{Results for 2D reconstruction guided vs unguided \textbf{with lesion} for noise level \textbf{2.5}. \blue{The points represent different values of the parameter $\lambda$, and the notation \protect \usym{2719} and $\Diamondblack$ denote the smallest and largest numerical value of $\lambda$, respectively.}}
    \label{fig:results_tumour_25_guided_sweep}
\end{figure}

\begin{figure}[h!]
    \vspace{-1.5em}
    \centering

\includegraphics[width=1.\linewidth,trim={0 0 0 0},clip]{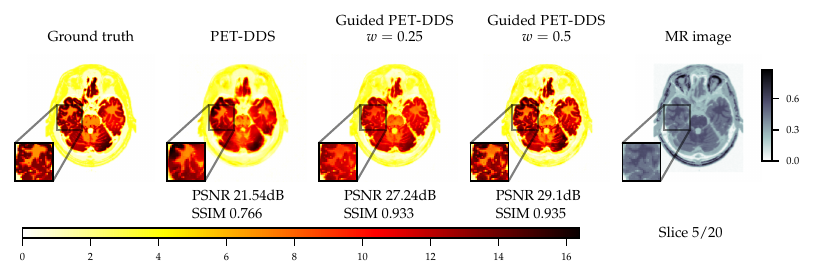} \\

 \vspace{-0.5em}
 
\includegraphics[width=1.\linewidth,trim={0 0 0 2.1em},clip]{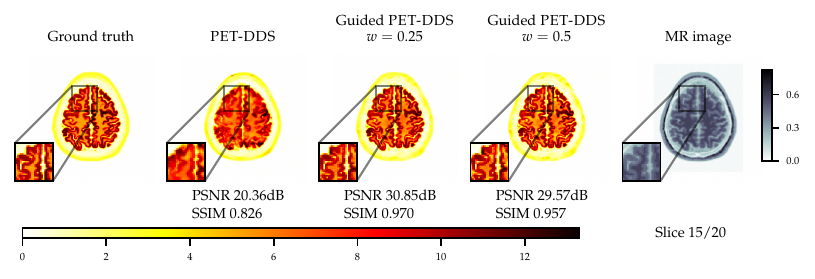}
  \caption{\vspace{-1em}Comparisons of the \PETDDS~MR guided vs. unguided at noise level \textbf{2.5 without lesions}.}
    \label{fig:2D_guided_img_ext_1}
    \vspace{-1em}
\end{figure}

\begin{figure}[h!]
    \vspace{-.5em}
    \centering
  \includegraphics[width=1.\linewidth,trim={0 0 0 0em},clip]{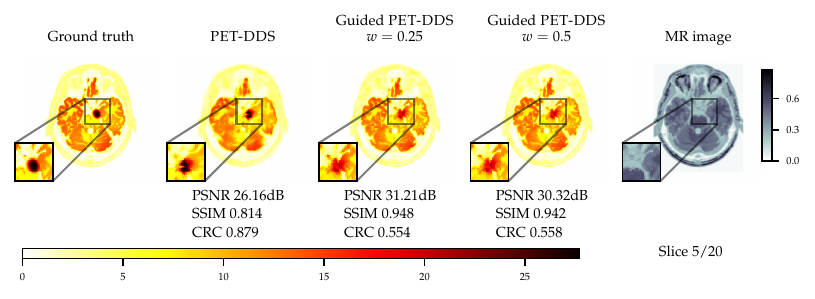}\\
  \vspace{-0.5em}
  \includegraphics[width=1.\linewidth,trim={0 0 0 2.1em},clip]{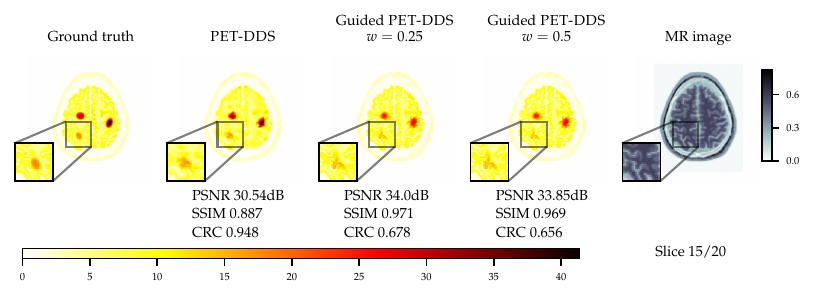}
    \caption{\vspace{-1em}Comparison of the \PETDDS~MR guided vs. unguided with a noise level \textbf{2.5} \textbf{with lesions}.}
    \label{fig:2D_guided_img_ext_2}
    \vspace{-1em}
\end{figure}

We show additional results for the MR guided model. Sensitivity plots without and with lesions are presented in Fig. \ref{fig:results_no_tumour_25_guided_sweep} and Fig. \ref{fig:results_tumour_25_guided_sweep}. These results support the findings of the paper, as the MR guided models achieve better reconstruction quality w.r.t. PSNR and SSIM. However, the CRC is similar to the unguided model. As the lesions were not visible in the MR image, no additional information about the lesions are introduced through guidance. We show two more reconstruction examples without lesions in Fig. \ref{fig:2D_guided_img_ext_1} and examples with lesions in Fig. \ref{fig:2D_guided_img_ext_2}.

\subsection{3D results RDPz sweeps}

\begin{figure}[h!]
    \centering
    \includegraphics[width=1.\textwidth]{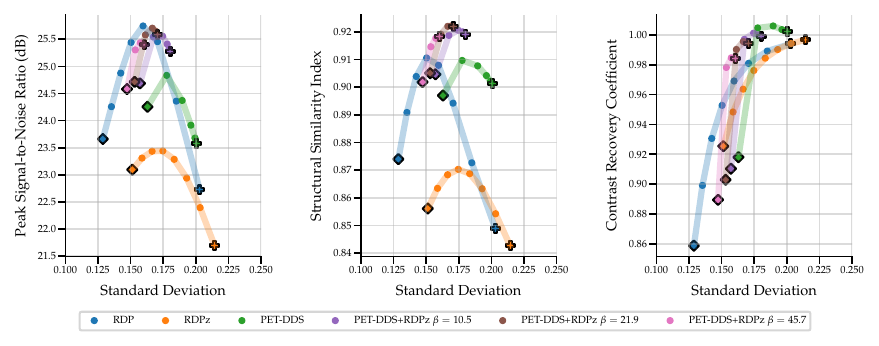}
    \caption{Results for 3D reconstruction using the FDG tracer for different penalty values. \blue{The points represent different values of the parameter $\lambda$, and the notation \protect \usym{2719} and $\Diamondblack$ denote the smallest and largest numerical value of $\lambda$, respectively.}}
    \label{fig:results_3d_fdg_rdpz_sweep}
\end{figure}

\begin{figure}[h!]
    \centering
    \includegraphics[width=1.\textwidth]{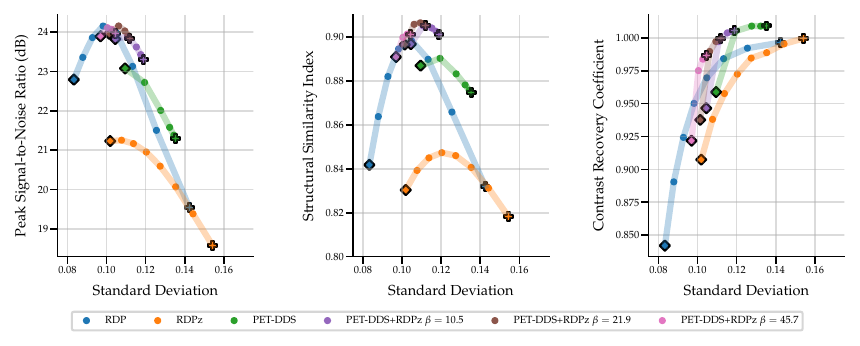}
    \caption{Results for 3D reconstruction using the Amyloid tracer for different penalty values. \blue{The points represent different values of the parameter $\lambda$, and the notation \protect \usym{2719} and $\Diamondblack$ denote the smallest and largest numerical value of $\lambda$, respectively.}}
    \label{fig:results_3d_amyloid_rdpz_sweep}
\end{figure}

\begin{figure}[h!]
    \centering
    \includegraphics[width=1\textwidth]{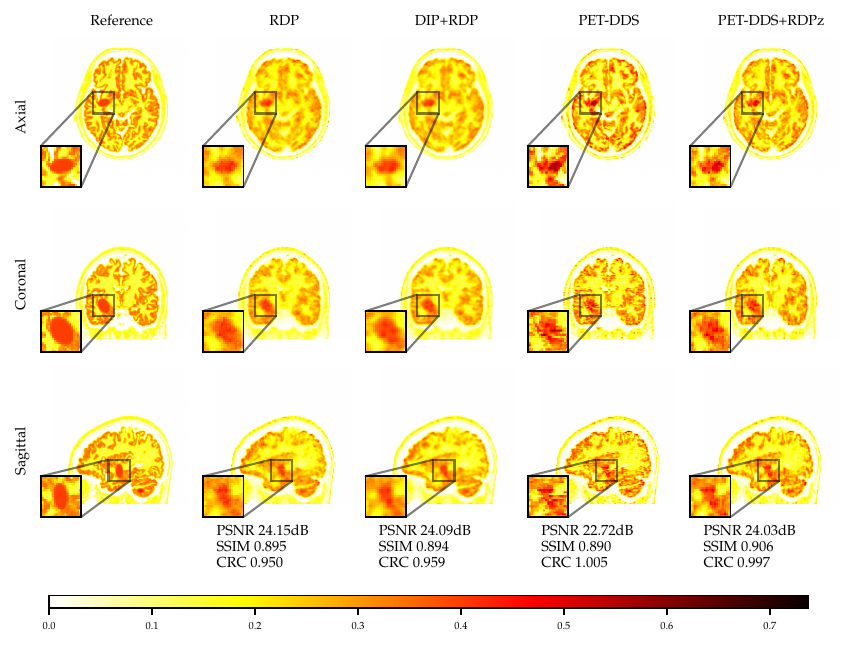}
    \caption{3D reconstruction for the different method with Amyloid tracer, and metrics computed on inset lesion.}
    \label{fig:results_3d_amyloid_images}
\end{figure}

We show the sensitivity plots for different penalty values of the additional RDP regularizer in $z$-direction for \PETDDS{} in Fig. \ref{fig:results_3d_fdg_rdpz_sweep} and \ref{fig:results_3d_amyloid_rdpz_sweep} for the two different tracers. In addition, we show axial, coronal and saggital slices of the reconstruction with the Amyloid tracer in \ref{fig:results_3d_amyloid_images}.

\begin{algorithm}[h!]
\caption{\PETNaive{}}\label{alg:nav}
\begin{algorithmic}
\Require Measurements $\By$
\Require Number of steps $N \in \N$
\Require Time discretisation $0=t_{k_1} \le \dots \le t_{k_N} = 1$
\State $\Bx_{t_N} \sim p_1$ \Comment{Sample initial noise}
\For {$k = N-1, \dots, 1$} 
\State $s_\theta \gets s_\theta(\Bx_{t_{k+1}}, t_{k+1})$
\State $\Bz \sim \mathcal{N}(0, \mathbf{I})$
\State $\Delta t \gets t_{k} - t_{k+1}$ 
\State $\Tilde{\Bx}_{t_{k}} \gets \Bx_{t_{k+1}} + \big[\Bf(\Bx_{t_{k+1}}, t_{k+1}) - g(t_{k+1})^2 s_\theta \big] \Delta t  + g(t_{k+1}) \sqrt{|\Delta t|} \Bz$ \Comment{Unconditional score update}
\State $\Bx_{t_{k}} \gets \tilde \Bx_{t_{k}}  - g(t_{k+1})^2 \lambda_{t_{k+1}}^\text{\Naive{}} \nabla_{\Bx} L(\By | c_{\text{OSEM}} P_{\Bx \geq 0}[\Bx_{t_{k+1}}]) \Delta t$ \Comment{Data consistency step}
\EndFor
\State $\hat{\Bx}  \gets c_{\text{OSEM}}\Bx_{t_1}$
\end{algorithmic}
\end{algorithm}

\begin{algorithm}[h!]
\caption{\PETDPS{}}\label{alg:dps}
\begin{algorithmic}
\Require Measurements $\By$
\Require Number of steps $N \in \N$
\Require Time discretisation $0=t_{k_1} \le \dots \le t_{k_N} = 1$
\Require Transition density $p(\Bxt | \Bxzero) = \mathcal{N}(\Bxt; \gamma_t \Bxzero, \nu_t^2 \mathbf{I})$
\State $\Bx_{t_N} \sim p_1$ \Comment{Sample initial noise}
\For {$k = N-1, \dots, 1$} 
\State $s \gets s_\theta(\Bx_{t_{k+1}}, t_{k+1})$
\State $\hat{\Bxzero}(\Bx_{t_{k+1}}) \gets \gamma_{t_{k+1}}^{-1} (\Bx_{t_{k+1}} + \nu_{t_{k+1}}^2 s_\theta) $ \Comment{Compute Tweedie estimate}
\State $\Bz \sim \mathcal{N}(0, \mathbf{I})$
\State $\Delta t \gets t_{k} - t_{k+1}$ 
\State $\Tilde{\Bx}_{t_{k}} \gets \Bx_{t_{k+1}} + \big[\Bf(\Bx_{t_{k+1}}, t_{k+1}) - g(t_{k+1})^2 s_\theta \big] \Delta t  + g(t_{k+1}) \sqrt{|\Delta t|} \Bz$ \Comment{Unconditional score update}
\State $\ell \gets L(\By | c_{\text{OSEM}} P_{\Bx \geq 0}[\hat{\Bxzero}(\Bx_{t_{k+1}})])$
\State $\Bx_{t_{k}} \gets \tilde \Bx_{t_{k}}  -  \lambda_{t_{k+1}}^\text{\texttt{DPS}}/\ell ~ \nabla_{\Bx} L(\By | c_{\text{OSEM}} P_{\Bx \geq 0}[\hat{\Bxzero}(\Bx_{t_{k+1}})]) $ \Comment{Data consistency step}
\EndFor
\State $\hat{\Bx} \gets c_{\text{OSEM}} \Bx_{t_1}$
\end{algorithmic}
\end{algorithm}

\begin{algorithm}[h!]
\caption{\PETDDS{}}\label{alg:dds}
\begin{algorithmic}
\Require Measurements $\By$
\Require Number of steps $N \in \N$
\Require Time discretisation $0=t_{k_1} \le \dots \le t_{k_N} = 1$
\Require Transition density $p(\Bxt | \Bxzero) = \mathcal{N}(\Bxt; \gamma_t \Bxzero, \nu_t^2 \mathbf{I})$
\Require Number of inner optimisation steps $p \in \N$, number of subsets $n_{\rm{sub}} \in \N$
\Require Stochasticity $\{\eta_t\}_{t\ge 0}$
\State $\Bx_{t_N} \sim p_1$ \Comment{Sample initial noise}
\For {$k = N-1, \dots, 1$} 
\State $s_\theta \gets s_\theta(\Bx_{t_{k+1}}, t_{k+1})$
\State $\hat{\Bxzero}(\Bx_{t_{k+1}})^0 \gets \gamma_{t_{k+1}}^{-1} (\Bx_{t_{k+1}} + \nu_{t_{k+1}}^2 s_\theta) $
\Comment{Compute Tweedie estimate}
\For {$i=0, \dots, p-1$} \Comment{Inner optimisation for data consistency}
\State $j \gets (p(N-k)+i \mod n_{\rm{sub}}) + 1$
\State $\Phi_{j}(\Bx^i_{t_{k+1}}) \gets  L_{j} (\By | c_\text{OSEM} \Bx^i_{t_{k+1}}) + (\lambda^\text{RDP}  R_z(\Bx^i_{t_{k+1}})- \lambda^\text{DDS}\| \Bx^i_{t_{k+1}} - \hat{\Bxzero}(\Bx_{t_{k+1}})^0 \|_2^2)/n_{\mathrm{sub}}$
\State $\Bx^{i+1}_{t_{k+1}} \gets P_{\mathbf{x}\geq 0} \left[  \Bx^{i}_{t_{k+1}} + \mathbf{D}(\Bx^{i}_{t_{k+1}})  \nabla_\Bx\Phi_{j}(\Bx^i_{t_{k+1}})\right]$ 
\EndFor
\State $\Bz \sim \mathcal{N}(0, \mathbf{I})$
\State $\text{Noise}(\Bx_{t_{k+1}}, s_\theta) \gets - \nu_{t_{k+1}} \sqrt{ \nu_{t_{k}}^2 - \eta_{t_{k+1}}^2 } s_\theta$
\State $\Bx_{t_{k}} \gets \gamma_{t_{k}} \Bx^{p}_{t_{k+1}} + \text{Noise}(\Bx_{t_{k+1}}, s_\theta) + \eta_{t_{k+1}} \Bz$
\EndFor
\State $\hat{\Bx} \gets c_{\text{OSEM}} \Bx_{t_1}$
\end{algorithmic}
\end{algorithm}

\end{document}

%% file: Macros.tex
\providecommand{\Bx}{{\mathbf{x}}}
\providecommand{\Tweedie}{{\hat{\mathbf{x}}_0}}

\providecommand{\Bxt}{{\mathbf{x}_t}}
\providecommand{\Bxzero}{{\mathbf{x}_0}}
\providecommand{\BxMR}{{\mathbf{x}_\text{MR}}}
\providecommand{\BxOSEM}{\mathbf{x}_\text{OSEM}}

\providecommand{\By}{{\mathbf{y}}}
\providecommand{\Bz}{{\mathbf{z}}}
\providecommand{\Bf}{{\mathbf{f}}}

\providecommand{\Bf}{{\mathbf{f}}}

\providecommand{\Bw}{{\mathbf{w}}}

\providecommand{\Bbarb}{\mathbf{\Bar{b}}}
\providecommand{\Bbary}{\mathbf{\Bar{y}}}

\providecommand{\BA}{{\mathbf{A}}}

\providecommand{\Naive}{\textsc{Naive}}

\providecommand{\LearnedPD}{\textsc{PET-LPD}}
\providecommand{\OSEMUNet}{\textsc{PET-UNet}}
\providecommand{\NaiveOSEM}{\textsc{Naive (OSEM)}}
\providecommand{\PETNaive}{\textsc{PET-Naive}}
\providecommand{\PETDPS}{\textsc{PET-DPS}}
\providecommand{\PETDDS}{\textsc{PET-DDS}}

\providecommand{\pPost}{{p^\text{post}}}
\providecommand{\pLkhd}{{p^\text{lkhd}}}
\providecommand{\pLkhdt}{{p_t}}
\providecommand{\pprior}{{\pi}}
\providecommand{\ppriort}{{p_t}}

\providecommand{\E}{{\mathbb E}}
\providecommand{\N}{{\mathbb N}}
\providecommand{\R}{{\mathbb R}}
\providecommand{\cP}{{\text{Pois}}}

\providecommand{\betamin}{\beta_\text{min}}
\providecommand{\betamax}{\beta_\text{max}}

